\documentclass[11pt]{article}	
\pdfoutput=1 
\usepackage{cite}
\usepackage{color} 
\usepackage{amsmath}
\usepackage{amsfonts}
\usepackage{amssymb}
\usepackage{array,booktabs}
\usepackage{placeins}
\usepackage{multirow}
\usepackage{float}
\usepackage{caption}
\usepackage{subcaption}
\usepackage{cancel}
\usepackage{graphicx}
\usepackage{slashed}            
\usepackage{bbm}                
\usepackage{xspace}				
\usepackage{collref}				
\usepackage{framed}				
\usepackage{placeins}
\usepackage{jheppub}							
\usepackage{mathrsfs}
\usepackage{hyperref}
\usepackage{tikz}
\usetikzlibrary{arrows,shapes}
\usetikzlibrary{trees}
\usetikzlibrary{matrix,arrows} 				
\usetikzlibrary{positioning}				
\usetikzlibrary{calc,through}				
\usetikzlibrary{decorations.pathreplacing}  
\usepackage{pgffor}							

\usetikzlibrary{decorations.pathmorphing}	
\usetikzlibrary{decorations.markings}
\usetikzlibrary{snakes}
\usepackage[normalem]{ulem}

\tikzset{
    vector/.style={decorate, decoration={snake}, draw},
	provector/.style={decorate, decoration={snake,amplitude=2.5pt}, draw},
	antivector/.style={decorate, decoration={snake,amplitude=-2.5pt}, draw},
    fermion/.style={draw, postaction={decorate},
        decoration={markings,mark=at position .55 with {\arrow[draw]{>}}}},
    fermionbar/.style={draw, postaction={decorate},
        decoration={markings,mark=at position .55 with {\arrow[draw=black]{<}}}},
    fermionnoarrow/.style={draw},
    gluon/.style={decorate, draw,decoration={coil,amplitude=4pt, segment length=6pt}, line width=1},
    scalar/.style={dashed,draw, postaction={decorate},
        decoration={markings,mark=at position .55 with {\arrow[draw]{>}}}},
    scalarbar/.style={dashed,draw, postaction={decorate},
        decoration={markings,mark=at position .55 with {\arrow[draw]{<}}}},
    scalarnoarrow/.style={dash pattern = on 6 pt off 3 pt,draw},
    electron/.style={draw, postaction={decorate},
        decoration={markings,mark=at position .55 with {\arrow[draw]{>}}}},
	bigvector/.style={decorate, decoration={snake,amplitude=4pt}, draw},
	vectorscalar/.style={loosely dotted,draw, postaction={decorate}},
}
\usepackage{relsize}
\usepackage{slashed}

\def\beq{\begin{equation}}
\def\eeq{\end{equation}}

\def\bea{\begin{eqnarray}}
\def\eea{\end{eqnarray}}
\def\bei{\begin{itemize}}
\def\eei{\end{itemize}}
\def\={\,=\,}
\def\+{\,+\,}
\def\-{\,-\,}
\def\GeV{\,{\rm GeV}\,}

\newcommand{\Eq}[1]{Eq.~(\ref{#1})}

\begin{document}

\def\lsim{\mathrel{\rlap{\lower4pt\hbox{\hskip1pt$\sim$}}
    \raise1pt\hbox{$<$}}}
\def\gsim{\mathrel{\rlap{\lower4pt\hbox{\hskip1pt$\sim$}}
    \raise1pt\hbox{$>$}}}
\newcommand{\vev}[1]{ \left\langle {#1} \right\rangle }
\newcommand{\bra}[1]{ \langle {#1} | }
\newcommand{\ket}[1]{ | {#1} \rangle }
\newcommand{\ev}{ {\rm eV} }
\newcommand{\kev}{{\rm keV}}
\newcommand{\mev}{{\rm MeV}}
\newcommand{\gev}{{\rm GeV}}
\newcommand{\tev}{{\rm TeV}}
\newcommand{\mpl}{$M_{Pl}$}
\newcommand{\mw}{$M_{W}$}
\newcommand{\Ft}{F_{T}}
\newcommand{\Zparity}{\mathbb{Z}_2}
\newcommand{\BLambda}{\boldsymbol{\lambda}}
\newcommand{\be}{\begin{eqnarray}}
\newcommand{\ee}{\end{eqnarray}}
\newcommand{\met}{\;\not\!\!\!{E}_T}

\newcommand{\sla}[1]{\setbox0=\hbox{$#1$}           
   \dimen0=\wd0                                     
   \setbox1=\hbox{/} \dimen1=\wd1                   
   \ifdim\dimen0>\dimen1                            
      \rlap{\hbox to \dimen0{\hfil/\hfil}}          
      #1                                            
   \else                                            
      \rlap{\hbox to \dimen1{\hfil$#1$\hfil}}       
      /                                             
   \fi} 

\newcommand{\Tab}[1]{Table~\ref{tab:#1}}
\newcommand{\tab}[1]{table~\ref{tab:#1}}
\newcommand{\tabl}[1]{\label{tab:#1}}
\newcommand{\Fig}[1]{figure~\ref{fig:#1}}
\newcommand{\Figl}[1]{\label{fig:#1}}
\newcommand{\draftnote}[1]{{\bf
 #1}}
\def\YT#1{{\bf  \textcolor{red}{[YT: {#1}]}}}
\def\ES#1{{\bf  \textcolor{blue}{[ES: {#1}]}}}
\def\HC#1{{\bf  \textcolor{orange}{[HC: {#1}]}}}
\def\SJ#1{{\bf  \textcolor{green}{[SJ: {#1}]}}}

\title{Exotic Quarks in Twin Higgs Models}
\author{Hsin-Chia Cheng,$^1$ Sunghoon Jung,$^{2,3}$ Ennio Salvioni,$^{1}$ and Yuhsin Tsai$^{1,4}$}
\affiliation{$^1$Department of Physics, University of California, Davis, Davis, CA 95616, USA\\$ ^2$Korea Institute for Advanced Study, Seoul 130-722, Korea \\$^3$SLAC National Accelerator Laboratory, 2575 Sand Hill Road, Menlo Park, CA 94025, USA \\$^4$Maryland Center for Fundamental Physics, Department of Physics, University of Maryland, College Park, MD 20742, USA}
\emailAdd{cheng@physics.ucdavis.edu}
\emailAdd{shjung@slac.stanford.edu}
\emailAdd{esalvioni@ucdavis.edu}
\emailAdd{yhtsai@umd.edu}

\abstract{The Twin Higgs model provides a natural theory for the electroweak symmetry breaking without the need of new particles carrying the standard model gauge charges below a few TeV. In the low energy theory, the only probe comes from the mixing of the Higgs fields in the standard model and twin sectors. However, an ultraviolet completion is required below $\sim$ 10 TeV to remove residual logarithmic divergences. In non-supersymmetric completions, new exotic fermions charged under both the standard model and twin gauge symmetries have to be present to accompany the top quark, thus providing a high energy probe of the model. Some of them carry standard model color, and may therefore be copiously produced at current or future hadron colliders. Once produced, these exotic quarks can decay into a top together with twin sector particles. If the twin sector particles escape the detection, we have the irreducible stop-like signals. On the other hand, some twin sector particles may decay back into the standard model particles with long lifetimes, giving spectacular displaced vertex signals in combination with the prompt top quarks. This happens in the Fraternal Twin Higgs scenario with typical parameters, and sometimes is even necessary for cosmological reasons. We study the potential displaced vertex signals from the decays of the twin bottomonia, twin glueballs, and twin leptons in the Fraternal Twin Higgs scenario. Depending on the details of the twin sector, the exotic quarks may be probed up to $\sim$ 2.5 TeV at the LHC and beyond 10 TeV at a future 100 TeV collider, providing a strong test of this class of ultraviolet completions.

}  
\preprint{SLAC-PUB-16433, KIAS-P15042, UMD-PP-015-016}

\maketitle

\section{Introduction}
The discovery of the Higgs boson at the Run 1 of the Large Hadron Collider (LHC) marks the great success of the Standard Model (SM). On the other hand, the absence of evidence of new physics at the LHC so far also effectuates a big puzzle for physicists.  Since the Higgs is a scalar field, its mass is sensitive to physics at high energies. The quantum effects from the SM interactions of the Higgs field generate quadratically divergent corrections to the squared mass of the Higgs. The largest contribution comes from the coupling to the top quark. Unless it is cut off at a scale much beneath a TeV, it would introduce a naturalness problem for the weak scale, which is governed by the Higgs mass. Most new physics theories which solve the naturalness problem, such as weak-scale supersymmetry (SUSY) and composite Higgs, require at least a relatively light colored top partner to cancel the contribution from the SM top quark. Under typical assumptions, the search limits of these top partners at the LHC have reached 700--800 GeV, which already implies a $\sim$ 10\% tuning or worse. Although there are still holes in the search coverage and there are ways to hide these particles  in more sophisticated models, one may want to wonder about the possibility that there is no new light colored particle more seriously.

The Twin Higgs model was proposed by Chacko, Goh and Harnik in 2006~\cite{Chacko:2005pe} to demonstrate that it is possible to have a natural electroweak scale without new particles charged under the SM gauge symmetry below a few TeV. It is based on an (approximate) $Z_2$ symmetry which relates the SM sector and a mirror, or twin, sector. The mass term of the Higgs fields in these two sectors exhibits an enhanced $U(4)$ symmetry due to this $Z_2$ symmetry. When the combined Higgs multiplet gets a nonzero vacuum expectation value (VEV), it breaks $U(4)$ down to $U(3)$. To obtain a realistic model, we need the symmetry breaking VEV $f$ to point mostly in the twin sector direction, with a small component $v\simeq 246$~GeV in the SM sector to break the SM electroweak symmetry. This requires a soft breaking of the $Z_2$ symmetry. The SM Higgs field can be viewed as the pseudo-Nambu-Goldstone bosons of the symmetry breaking in the limit $v \ll f$. Among the 7 Nambu-Goldstone modes, 3 are eaten to become the longitudinal modes of the $SU(2)$ gauge bosons $W_B, \, Z_B$ of the twin sector, 3 are eaten to become the longitudinal modes of the SM $W$ and $Z$ bosons, and the last one corresponds to the Higgs boson of 125 GeV that we observed. The one-loop quadratically divergent contributions to the Higgs mass-squared from the SM particles are canceled by the one-loop contributions from the corresponding particles in the twin sector. The particles in the twin sector generally are heavier than the corresponding SM particles by a factor $f/v$ (if the their couplings to Higgs respect the $Z_2$ symmetry). To avoid an excessive fine-tuning (which is measured by $v^2/f^2$) we should have $f/v \lsim 5$. The cutoff of the theory can be raised to the scale $\Lambda_{\rm UV} \approx 4\pi f \sim $ 5--10 TeV. The twin sector particles do not carry SM gauge quantum numbers and hence are difficult to look for at colliders.

If the twin sector is an exact mirror of the SM sector except for the overall scale ratio $f/v$, there will be many light states in the twin sector, which could cause cosmological problems. In addition, the two ``photons'' in the two sectors in general will have a kinetic mixing, which is strongly constrained if the twin photon is massless or very light. On the other hand, these light particles play little role in stabilizing the Higgs mass as their couplings to the Higgs are small. Their existence is not necessary for the naturalness as long as the $Z_2$ symmetry among  the large couplings to the Higgs from the heavy states can be approximately preserved. In view of this, Craig, Katz, Strassler and Sundrum proposed a minimal model~\cite{Craig:2015pha}, dubbed ``Fraternal Twin Higgs,'' in which only the minimal particle content in the twin sector and approximate equality of the couplings for preserving the naturalness are kept. The necessary ingredients for the Fraternal Twin Higgs model were summarized in their paper as follows~\cite{Craig:2015pha}:
\begin{enumerate}
\item An additional twin Higgs doublet and an approximately $SU(4)$-symmetric potential.
\item Twin top and a twin top Yukawa that is numerically very close to the SM top Yukawa.
\item Twin weak bosons from the gauged $SU(2)$ with $\hat{g}_2(\Lambda_{\rm UV}) \approx g_2(\Lambda_{\rm UV})$.
\item Twin glue, a gauged $SU(3)$ symmetry with $\hat{g}_3(\Lambda_{\rm UV}) \approx g_3(\Lambda_{\rm UV})$. 
\item Twin bottom and twin tau, whose masses are essentially free parameters so long as they remain much lighter than the twin top.
\item Twin neutrino from the twin tau doublet, which may have a Majorana mass, again a free parameter as long as it is sufficiently light.
\end{enumerate}

In the low energy theory, the only connection between the SM sector and the twin sector is through the Higgs field. Due to the pseudo-Goldstone nature, the physical light Higgs boson does not align exactly in the direction of electroweak breaking VEV $v$, but has a mixture of the component in the twin sector direction with a mixing angle $\sim v/f$. The couplings of the Higgs boson to SM fields receive a universal suppression $\sim \cos\, (v/f)$. In addition, there are couplings (suppressed by $v/f$) between the Higgs boson and the twin sector fields so the Higgs boson can decay into the twin sector particles if the decay channels are open. In the Fraternal Twin Higgs model, the Higgs boson can decay into twin gluons, twin $b$'s, twin taus and twin neutrinos if they are light enough. Due to the strong twin color force, the twin gluons and twin bottoms will confine into twin glueballs and twin bottomonia. References \cite{Craig:2015pha,Curtin:2015fna} also studied their phenomenology. Among the stable twin hadrons, the $0^{++}$ glueball and bottomonium can mix with the Higgs boson and decay back to the SM particles~\cite{Juknevich:2009ji,Juknevich:2009gg}. Depending on the parameters, the lifetime is often quite long and can give rise to displaced decays at colliders, resulting in interesting signatures. If the twin sector particles do not decay back to SM particles inside the detector, one then has to rely on the invisible Higgs decay or precise measurement of the Higgs couplings to probe the model. However, these effects get diminished as the symmetry breaking scale $f$ is raised higher and the LHC reach is limited to $f/v \lsim 3$-$4$ \cite{Burdman:2014zta}.

The Twin Higgs model, as a low energy theory, requires an ultraviolet (UV) completion. Although the one-loop quadratically divergent contributions to the Higgs potential are canceled, there are still logarithmically divergent contributions. In particular the Higgs quartic coupling, which controls the physical Higgs boson mass, still receives divergent contributions because other states do not form complete $SU(4)$ representations. Likewise, some precision electroweak observables also receive divergent contributions due to the modification of the couplings.\footnote{We thank Roberto Contino for discussion on this point.} Therefore, in a UV-complete theory new states must appear at high energies to cut off these divergences. These new states are expected to appear below the scale $4\pi f$ where the theory becomes strongly coupled. A simple extension was already considered in the original Twin Higgs paper~\cite{Chacko:2005pe}. It involves embedding the top quarks of the SM and twin sectors into complete $SU(6)\times SU(4)\, [\supset (SU(3)\times SU(2))^2]$ multiplets. This requires new fermions which are charged under both the SM and the twin gauge group. Such new fermions also appear in all non-supersymmetric UV completions of the Twin Higgs model that have been constructed so far.\footnote{In supersymmetric UV completions~\cite{Falkowski:2006qq,Chang:2006ra,Craig:2013fga} there are usual colored top squarks (stops) which can be searched for, albeit being heavier than in the standard SUSY scenario. Reference \cite{Batra:2008jy} studied a composite Left-Right Twin Higgs model, which is different from what we consider here.} For composite UV completions~\cite{Barbieri:2015lqa,Low:2015nqa}, these new fermions are resonances of the composite dynamics, and in UV completions with extra dimensions~\cite{Geller:2014kta,Craig:2014aea,Craig:2014roa}, they are Kaluza-Klein (KK) excitations of bulk fields whose zero modes are removed by boundary conditions or orbifold projections. These new fermions connect the SM sector and the twin sector and provide another interesting probe of the Twin Higgs model. In particular, the fermions that carry SM color quantum numbers (which we call exotic quarks) may have significant production cross sections if they are light enough to be within the reach of LHC or a future higher energy collider. 

In this paper, we study the collider phenomenology associated with these exotic quarks with the assumption that the low energy theory is essentially described by  the Fraternal Twin Higgs model (with possible small variations). These exotic quarks also carry the twin $SU(2)$ quantum numbers. They can decay into SM top quarks plus $W_B,\, Z_B$ gauge bosons of the twin sector. The twin gauge bosons further decay into twin fermions.  The $Z_B$ decay into a twin $b$ pair can give rise to interesting quirky behavior~\cite{Okun:1980mu,Bjorken,Gupta:1981ve,Kang:2008ea} depending on the twin $b$ mass. In general multiple twin glueballs and/or twin bottomonia will be produced at the end. Some of them can have displaced decays back into SM particles  by mixing with the Higgs or through the kinetic mixing between the SM and twin $U(1)$ gauge bosons, if the latter exists. These displaced signals are easy to trigger on, due to the presence of the additional prompt top quarks in the events. The search reach of the exotic quarks extends above $\sim$ 2 TeV at the LHC and $\sim$ 10 TeV at a future 100 TeV collider, often besting the standard stop or $t'$ searches. The twin tau and twin neutrino are singlets under both the SM gauge group and the unbroken twin gauge group, so they act like sterile neutrinos to us. They could couple to SM leptons through higher dimensional operators. In that case, they can also give interesting displaced vertex signals from their decays into SM quarks or leptons, albeit more model-dependent.

This paper is organized as follows. In Sec.~\ref{sec:model} we briefly review the Twin Higgs model and point out the necessity of the exotic fermions for the UV completion. We also discuss the relation of their masses and the Higgs boson mass. In Sec.~\ref{sec:production} we derive the interactions of the exotic quarks and calculate their production cross sections at hadron colliders and their decay branching ratios. We also obtain bounds and future experimental reaches on their masses by scaling up the current fermionic top partner searches and stop searches. Section~\ref{sec:hiddenpheno} contains our main results. We consider the collider phenomenology when the exotic quarks decay into the twin sector, and the produced twin sector states decay back to the SM particles with displaced vertices. The three subsections discuss the twin bottomonium, twin glueball, and twin lepton cases, respectively. The conclusions are drawn in Sec.~\ref{sec:conclusions}, where we also discuss how variations of our benchmark model may affect the signals. In App.~\ref{app:multiplicity} we detail our estimates of the number of twin hadrons produced in the exotic quark decay, using a simplified string model. In App.~\ref{app:bottomonium} we give our calculations of the bottomonium decay rates, and verify them against the experimental data of the SM meson decays.

\section{The Twin Higgs Model}
\label{sec:model}

We start by a brief description of the Twin Higgs mechanism. One assumes that there is a twin sector of the SM sector and it is related to the SM by an approximate $Z_2$ symmetry. Due to the $Z_2$ symmetry, the quadratic term of the Higgs potential of the twin sector and of the SM preserves an enhanced $U(4)$ symmetry.  The global $U(4)$ is spontaneously broken down to $U(3)$ by the Higgs VEV, giving rise to 7 Nambu-Goldstone modes. We shall assume that the radial mode is heavy, not present below the cutoff, so that $U(4)$ is nonlinearly realized below the cutoff. The Higgs is represented by the nonlinear field,
\begin{equation}
H= \begin{pmatrix} H_B \\ H_A \end{pmatrix}  \equiv \exp\left(i \frac{\Pi}{f}\right)\begin{pmatrix} f/\sqrt{2} \\ 0 \\ 0 \\ 0 \end{pmatrix},
\end{equation}
where the subscripts $A$ and $B$ represent the SM sector and the twin sector respectively. (We arrange the twin sector in front for later convenience.) The Nambu-Goldstone field matrix $\Pi$ is given by
\begin{align}
\Pi 
= \begin{pmatrix} \pi_7& \pi_5+i\pi_6 & \pi_3+i\pi_4 & \pi_1 + i\pi_2 \\
\pi_5-i\pi_6 & &  & \\
\pi_3 - i\pi_4 & & \mathbf{0}_3 & \\
\pi_1 - i\pi_2 & & & \end{pmatrix} \, .
\end{align} 
Expanding the exponential we find the explicit form
\begin{align}
H= \begin{pmatrix} H_B \\ H_A \end{pmatrix} \,=&\, \frac{f}{\sqrt{2}\tilde{\pi}}\exp\left(\frac{i\pi_7}{2 f}\right)\begin{pmatrix} \tilde{\pi}\cos(\tilde{\pi}/f) + i(\pi_7/2)\sin(\tilde{\pi}/f) \\ \sin (\tilde{\pi}/f)\begin{pmatrix}  i \pi_5 + \pi_6 \\ i \pi_3 + \pi_4 \\
i \pi_1 + \pi_2 \end{pmatrix}  \end{pmatrix}  \nonumber\\ 
\,\simeq &\, \frac{1}{\sqrt{2}}\begin{pmatrix} f +  i \pi_7 \\ i \pi_5 + \pi_6 \\ i \pi_3 + \pi_4 \\ i \pi_1 + \pi_2 \end{pmatrix} + O(1/f) \label{expansion}
\end{align}
where we defined $\tilde{\pi}\equiv \sqrt{\pi_1^2 + \ldots + \pi_6^2 + \pi_7^2/4}\,$.

The SM electroweak gauge group $SU(2)_A \times U(1)_Y$ and its mirror gauge group $SU(2)_B \times U(1)_D$ in the twin sector\footnote{Naturalness does not require gauging $U(1)_D$. In the minimal Fraternal Twin Higgs model it is therefore assumed that $U(1)_D$ is not gauged. We will still consider that $U(1)_D$ is gauged but the gauge symmetry is broken so that its gauge boson acquires a mass. The ungauged case corresponds to the decoupling limit where the gauge boson mass is taken to infinity.} are embedded in the $U(4)= SU(4) \times U(1)_X$ in the Higgs sector as follows,
\begin{align}
&SU(4) \supset  \begin{pmatrix} SU(2)_B & \\
& SU(2)_A \end{pmatrix}\, , \\
&Y= -\frac{1}{2} \begin{pmatrix} \mathbf{0}_2 & \\
& \mathbf{1}_2\end{pmatrix} = -\frac{1}{4} + \frac{1}{4} \begin{pmatrix} \mathbf{1}_2 & \\
& -\mathbf{1}_2\end{pmatrix} =X + \frac{1}{\sqrt{2}} T_d^{SU(4)}\, ,\label{eq:abelian1} \\
&D= -\frac{1}{2} \begin{pmatrix} \mathbf{1}_2 & \\
& \mathbf{0}_2\end{pmatrix}  =-\frac{1}{4}  - \frac{1}{4} \begin{pmatrix} \mathbf{1}_2 & \\
& -\mathbf{1}_2\end{pmatrix}=X - \frac{1}{\sqrt{2}} T_d^{SU(4)}\, .
\label{eq:abelian2}
\end{align}
The $U(1)_Y$ and $U(1)_D$ are linear combinations of $U(1)_X$ (with $X=-1/4$ for $H$) and the diagonal $U(1)$ subgroup of $SU(4)$ with  the generator $T_d^{SU(4)}$ proportional to ${\rm diag} (\mathbf{1}_2 \,, 
 -\mathbf{1}_2)$. Their normalizations are chosen such that the SM Higgs field has hypercharge $Y=-1/2$ and no $D$ charge, and vice versa for the twin sector Higgs.\footnote{The choice of $Y=-1/2$ for the Higgs is more convenient later for writing down the top Yukawa coupling.} The two $SU(2)$ gauge couplings need to be approximately equal by the $Z_2$ symmetry to preserve the $U(4)$ symmetry of the Higgs mass term. On the other hand, the $U(1)$ couplings do not need to be related as long as they are small enough not to affect the naturalness.
 
Six of the seven Goldstone bosons are eaten by the $W, \, Z$ bosons of the SM sector and twin sector. Going to the unitary gauge $\pi_4 = h$, $\pi_i = 0$ for $i\neq 4$, we find
\begin{equation} \label{UG}
H\to \begin{pmatrix} \frac{f}{\sqrt{2}}\cos\frac{h}{f} \\ 0 \\ \frac{f}{\sqrt{2}} \sin\frac{h}{f} \\0 \end{pmatrix}.
\end{equation}
To have a viable model we require $\left\langle h \right\rangle \ll f$. This can be achieved by adding a soft $Z_2$ breaking mass term
\begin{equation}\label{softZ2break}
\mu^2 H_A^\dagger H_A \, ,
\end{equation}
which suppresses the $H_A$ VEV relative to the $H_B$ VEV.
Expanding the $H$ kinetic term $(D_\mu H)^\dagger D^\mu H$ we obtain the masses of $W_{A,B}$,
\begin{equation}
m_{W_A}^2 = \frac{g^2 f^2}{4}\sin^2\left(\frac{v_A}{f}\right) = \frac{g^2 v^2}{4}\,,\qquad m_{W_B}^2 = \frac{g^2 f^2}{4}\cos^2\left(\frac{v_A}{f}\right) = \frac{g^2 f^2}{4}\left(1-\frac{v^2}{f^2}\right),
\end{equation}
where $v_A \equiv \left\langle h \right\rangle$ and $v=f\sin\left(v_A/f\right)\simeq 246$~GeV. The masses of $Z_{A,B}$ are obtained by replacing $g\to g/\cos\theta_w$ where $\theta_w$ is the Weinberg angle.

%
%
In the fermion sector, the top Yukawa interaction and the corresponding term in the twin sector are given by
\begin{equation} \label{minimalYukawa}
-y_t H_A^\dagger \bar{u}_{3R}^A q_{3L}^A -\hat{y}_t H_B^\dagger \bar{u}_{3R}^B q_{3L}^B\,, 
\end{equation}
where $q_{3L}^B,\, u_{3R}^B$ are the twin top partners which are completely neutral under the SM gauge interactions. The approximate $Z_2$ which exchanges $A\leftrightarrow B$ requires $\hat{y}_t \approx y_t$ so that the radiative corrections to the Higgs mass term from the large top Yukawa interactions remains approximately $U(4)$ invariant. To avoid gauge anomalies in the twin sector, the mirror partners $d_{3R}^B$, $\ell_{3L}^B=(\tau_L^B, \nu_L^B)$, $\tau_R^B$ of the SM third generation fermions also need to be included. The Yukawa couplings of the twin $b$ and twin leptons are less constrained as long as they are much smaller than the top Yukawa coupling because their contributions to the Higgs mass are small. For the same reason, the mirror partners of the first two generation SM fermions are not needed for the naturalness of the weak scale. This is the point advocated by the Fraternal Twin Higgs model and will be adopted in this paper.

A further ingredient of the model is the gauging of twin color, with $\hat{g}_{s} \approx g_s$ at the cutoff scale. This prevents the running contribution due to $g_s$ from spoiling the approximate equality $y_t \approx \hat{y}_t$ in the infrared. In the case that the twin QCD has only two flavors as we assumed, its coupling runs faster, leading to a higher confinement scale $\Lambda$ compared to the SM QCD. If we take $\hat{g}_{s} = g_s$ at 5 TeV, we find a twin confinement scale $\Lambda\simeq 5.3$ GeV (including the two-loop corrections to the $\beta$ functions in the $\overline{\rm MS}$ scheme) for $f=1\;\mathrm{TeV}$. If we allow 10\% difference between $\hat{g}_s$ and $g_s$ (at 5 TeV), $\Lambda$ can vary roughly between 1 to 20 GeV~\cite{Craig:2015pha}.

From Eq.~\eqref{UG} we read that the couplings of the Higgs to $W,Z$ and to fermions are modified by the universal factor $\sqrt{1-v^2/f^2}\,$. Therefore the scale $f$ is already mildly constrained by Higgs coupling measurements at the $8$ TeV LHC, $v^2/f^2 \lesssim 0.2$. The projected LHC sensitivity is, however, limited, and $f\gtrsim 1\;\mathrm{TeV}$ would be out of reach even in the high-luminosity phase. In that case, tests of Twin Higgs at hadron colliders could only occur through direct signatures, such as those studied in this paper. On the other hand, since the Yukawa couplings of the twin $b$ and twin tau are not required to be equal to those of their SM counterparts, the invisible decay of the Higgs boson to twin $b$'s and twin taus can be enhanced if their Yukawa couplings are larger while the decays are still kinematically allowed. For example, for $f = 1\;\mathrm{TeV}$, the analysis of the current Higgs data gives a bound $\hat{y}_b/y_b < 2$ at $95\%$ CL~\cite{Craig:2015pha} or equivalently (in the rest of the paper we denote most of the twin sector mass eigenstates with a hat, except the gauge bosons $W_B,\, Z_B$)
\begin{equation}\label{eq:invHbound}
m_{\hat{b}} < 37\;\mathrm{GeV}\,\qquad \mathrm{or}\qquad m_{\hat{b}} > 62.5\;\mathrm{GeV}\,,
\end{equation}        
where the second inequality corresponds to the threshold for the $h\to \bar{\hat{b}} \hat{b}$ decay. 

Finally, the Higgs boson mass of 125~GeV also impose a non-trivial requirement on the Twin Higgs model. As we discussed earlier, the $Z_2$ symmetry ensures the $SU(4)$ invariance of the quadratic term of the Higgs potential. However it does not imply that the quartic and higher order terms are $SU(4)$ invariant.
In particular, the term
\begin{equation}
|H_A|^4+|H_B|^4
\end{equation}
is $Z_2$ symmetric but not $SU(4)$ invariant. Such explicit $SU(4)$ breaking terms inevitably exist because they will be generated by the interactions of the Higgs fields with the SM sector and the twin sector, and they are actually needed to give the physical Higgs boson a mass. The quartic term receives a logarithmic divergent contribution dominated by the top loop and the twin top loop, so it must be cut off by a counter term from the UV physics because its coefficient is related to the Higgs boson mass. The top and twin top contributions make the coefficient run towards negative in the UV. To avoid the appearance of other deeper vacua near the cutoff scale, which may destabilize our electroweak breaking vacuum, it is desirable to have the coefficient of the quartic term stay positive at the UV cutoff.

In a non-supersymmetric UV completion, to cut off the divergence from the top and twin top loops most likely requires states to fill complete multiplets of the $SU(4)$ symmetry in the UV.
This implies there will be new states charged under both the SM and twin gauge group. Indeed, in the original Twin Higgs paper~\cite{Chacko:2005pe}, Chacko, Goh, and Harnik already considered an extended fermion sector where new fermions together with the top and twin top quarks form complete multiplets of the enlarged global symmetry $SU(6)\times SU(4)\times U(1)$, in which the two copies of the subgroup $[SU(3)\times SU(2)\times U(1)]_{A,B}$ are gauged. They showed that the Higgs potential generated from this extended fermion sector is finite and calculable. Such new fermion states also exist in all non-supersymmetric UV completions of the Twin Higgs model so far, as the resonances of the composite dynamics~\cite{Barbieri:2015lqa,Low:2015nqa} or KK excitations in models with extra dimensions~\cite{Geller:2014kta,Craig:2014aea,Craig:2014roa}. The new fermions charged under both SM gauge group and the twin gauge group provide another bridge between the SM and twin sectors in addition to the Higgs. In particular, the new fermions carrying SM color (which we call exotic quarks) could be copiously produced if their masses are within the reach of the center of mass energies of the LHC or a future high energy collider. They also carry the electroweak charges of the twin gauge group and hence can decay to states in the twin sector. The purpose of this paper is to study the collider phenomenology of these exotic quarks. They provide another probe of the Twin Higgs model and also a direct test of its UV completion.

\subsection{Exotic Quarks}
Here we review the extended fermion sector in Ref.~\cite{Chacko:2005pe}. It can be considered as the lightest resonances or KK modes in a more UV-complete model. To make the contribution from the top sector to the Higgs potential finite, the global symmetry of the top Yukawa coupling is enlarged to $SU(6)\times SU(4)\times U(1)_X$, with the subgroup $SU(3)_B\times SU(3)_A\subset SU(6)$ gauged in addition to $SU(2)_B\times SU(2)_A\times U(1)_D\times U(1)_Y$. The definition of the gauged $U(1)$ generators in Eq.~\eqref{eq:abelian1},~\eqref{eq:abelian2} is extended to
\begin{align}
& Y = X + \frac{1}{\sqrt{2}}T_d^{SU(4)} - \frac{2}{\sqrt{3}}T_{d^\prime}^{SU(6)} = X + \frac{1}{4} \begin{pmatrix} \mathbf{1}_2  \\
& -\mathbf{1}_2\end{pmatrix}_{SU(4)}  -\frac{1}{3} \begin{pmatrix} \mathbf{1}_3 & \\ & -\mathbf{1}_{3}\end{pmatrix}_{SU(6)}\,,\\
& D = X - \frac{1}{\sqrt{2}}T_d^{SU(4)} + \frac{2}{\sqrt{3}}T_{d^\prime}^{SU(6)} = X- \frac{1}{4} \begin{pmatrix} \mathbf{1}_2  \\
& -\mathbf{1}_2\end{pmatrix}_{SU(4)}  +\frac{1}{3} \begin{pmatrix} \mathbf{1}_3 & \\ & -\mathbf{1}_{3}\end{pmatrix}_{SU(6)} \,, 
\end{align}
where
\begin{equation}
 T_{d^\prime}^{SU(6)} =\frac{1}{2\sqrt{3}}\begin{pmatrix} \mathbf{1}_3 & \\ & -\mathbf{1}_{3}\end{pmatrix}\,, 
\end{equation}
is the a diagonal generator of $SU(6)$ which commutes with $SU(3)_B$ and $SU(3)_A$. 
The fermion Yukawa coupling $\mathcal{L}_t$ and the mass term $\mathcal{L}_{m}$ are given by
\begin{equation}
-\mathcal{L}_t= y_t H^\dagger Q_{3L} \bar{u}_{3R} + \mathrm{h.c.} = y_t \begin{pmatrix} H_B^\dagger & H_A^\dagger\end{pmatrix}\begin{pmatrix} q^B_{3L} & \tilde{q}^A_{3L} \\ \tilde{q}^B_{3L} & q^A_{3L} \end{pmatrix} \begin{pmatrix} \overline{u}^B_{3R} \\ \overline{u}^A_{3R} \end{pmatrix} + \mathrm{h.c.}
\label{eq:yukawa}
\end{equation}
where under $SU(6)\times SU(4)\times U(1)_X$ we have $Q_{3L} \sim (\mathbf{6}, \mathbf{4}, 1/12)$, $u_{3R} \sim (\mathbf{6}, \mathbf{1}, 1/3)$ and $H\sim (\mathbf{1}, \mathbf{4}, -1/4)$, and 
\begin{equation}
-\mathcal{L}_{m} = \tilde{M}(\overline{\tilde{q}}^{A}_{3R} \tilde{q}_{3L}^A + \overline{\tilde{q}}^{B}_{3R} \tilde{q}_{3L}^B) + \mathrm{h.c.} \, .
\label{eq:mass}
\end{equation}
Because of mass mixings, these gauge eigenstates are not mass eigenstates. To distinguish them from the physical top quark which is a mass eigenstate, we use subscript ``3'' to denote the gauge eigenstates.
The fermion kinetic terms can be written in terms of the component fields given in Table~\ref{Tab:charges}. The states contained in $\tilde{q}_3^A$ and $\tilde{q}_3^B$ are charged under both sectors. In particular, $\tilde{q}_3^A$ is a $SU(2)_B$-doublet of fermions that carry SM color and are vector-like with respect to the SM electroweak interactions. These ``exotic quarks'' are the focus of our study. 
\begin{table}
   \begin{center}
   \begin{tabular}{c|c c|c c|c c|c c} 
    & \multicolumn{2}{c|}{$SU(3)$} & \multicolumn{2}{c|}{$SU(2)$} & \multicolumn{2}{c|}{$U(1)$} & \multicolumn{2}{c}{$U(1)_{\rm em}$} \\
    & $A$ & $B$ & $A$ & $B$ & $Y$ & $D$ & SM & Twin \\
   \hline
   &&&&&& \\[-0.4cm]
   $q^A_{3L} = \begin{pmatrix} u^A_{3L} \\ d^A_{3L} \end{pmatrix}$ & $\mathbf{3}$ & $\mathbf{1}$ & $\mathbf{2}$ & $\mathbf{1}$ & $1/6$ & $0$ & $\begin{pmatrix} 2/3 \\ -1/3 \end{pmatrix}$ & $\begin{pmatrix} 0 \\ 0 \end{pmatrix}$ \\[0.2cm]
   &&&&&& \\[-0.4cm]    
   $\tilde{q}^A_3 = \begin{pmatrix} \tilde{u}^A_3 \\ \tilde{d}^A_3 \end{pmatrix}$ & $\mathbf{3}$ & $\mathbf{1}$ & $\mathbf{1}$ & $\mathbf{2}$ & $2/3$ & $-1/2$ & $\begin{pmatrix} 2/3 \\ 2/3 \end{pmatrix}$ & $\begin{pmatrix} 0 \\ -1 \end{pmatrix}$ \\      
   &&&&&& \\[-0.4cm]
   $q^B_{3L} = \begin{pmatrix} u^B_{3L} \\ d^B_{3L} \end{pmatrix}$ & $\mathbf{1}$ & $\mathbf{3}$ & $\mathbf{1}$ & $\mathbf{2}$ & $0$ & $1/6$ & $\begin{pmatrix} 0 \\ 0 \end{pmatrix}$ & $\begin{pmatrix} 2/3 \\ -1/3 \end{pmatrix}$ \\
   &&&&&& \\[-0.4cm]
   $\tilde{q}^B_3 = \begin{pmatrix} \tilde{u}^B_3 \\ \tilde{d}^B_3 \end{pmatrix}$ & $\mathbf{1}$ & $\mathbf{3}$ & $\mathbf{2}$ & $\mathbf{1}$ & $-1/2$ & $2/3$ & $\begin{pmatrix} 0 \\ -1 \end{pmatrix}$ & $\begin{pmatrix} 2/3 \\ 2/3 \end{pmatrix}$ \\
   &&&&&& \\[-0.4cm]
   $u^A_{3R}$ & $\mathbf{3}$ & $\mathbf{1}$ & $\mathbf{1}$ & $\mathbf{1}$ & $2/3$ & $0$ & $2/3$ & $0$ \\
   &&&&&& \\[-0.4cm]
   $d^A_{3R}$ & $\mathbf{3}$ & $\mathbf{1}$ & $\mathbf{1}$ & $\mathbf{1}$ & $-1/3$ & $0$ & $-1/3$ & $0$ \\
   &&&&&& \\[-0.4cm]
   $u^B_{3R}$ & $\mathbf{1}$ & $\mathbf{3}$ & $\mathbf{1}$ & $\mathbf{1}$ & $0$ & $2/3$ & $0$ & $2/3$ \\
   &&&&&& \\[-0.4cm]
   $d^B_{3R}$ & $\mathbf{1}$ & $\mathbf{3}$ & $\mathbf{1}$ & $\mathbf{1}$ & $0$ & $-1/3$ & $0$ & $-1/3$ \\
   &&&&&& \\[-0.4cm]
   $\ell^A_{3L} = \begin{pmatrix} \nu_L^A \\ \tau_L^A \end{pmatrix}$ & $\mathbf{1}$ & $\mathbf{1}$ & $\mathbf{2}$ & $\mathbf{1}$ & $-1/2$ & $0$ & $\begin{pmatrix} 0 \\ -1 \end{pmatrix}$ & $\begin{pmatrix} 0 \\ 0 \end{pmatrix}$ \\
   &&&&&& \\[-0.4cm]
   $\ell^B_{3L} = \begin{pmatrix} \nu_L^B \\ \tau_L^B \end{pmatrix}$ & $\mathbf{1}$ & $\mathbf{1}$ & $\mathbf{1}$ & $\mathbf{2}$ & $0$ & $-1/2$ & $\begin{pmatrix} 0 \\ 0 \end{pmatrix}$ & $\begin{pmatrix} 0 \\ -1 \end{pmatrix}$ \\
   &&&&&& \\[-0.4cm]
   $\tau^A_{3R}$ & $\mathbf{1}$ & $\mathbf{1}$ & $\mathbf{1}$ & $\mathbf{1}$ & $-1$ & $0$ & $-1$ & $0$ \\
   &&&&&& \\[-0.4cm]
   $\tau^B_{3R}$ & $\mathbf{1}$ & $\mathbf{1}$ & $\mathbf{1}$ & $\mathbf{1}$ & $0$ & $-1$ & $0$ & $-1$ \\
   &&&&&& \\[-0.4cm]
   \hline
   &&&&&& \\[-0.4cm]
   $H_A$ & $\mathbf{1}$ & $\mathbf{1}$ & $\mathbf{2}$ & $\mathbf{1}$ & $-1/2$ & $0$ & $\begin{pmatrix} 0 \\ -1 \end{pmatrix}$ & $\begin{pmatrix} 0 \\ 0 \end{pmatrix}$ \\
   &&&&&& \\[-0.4cm]
   $H_B$ & $\mathbf{1}$ & $\mathbf{1}$ & $\mathbf{1}$ & $\mathbf{2}$ & $0$ & $-1/2$ & $\begin{pmatrix} 0 \\ 0 \end{pmatrix}$ & $\begin{pmatrix} 0 \\ -1 \end{pmatrix}$ \\
   \end{tabular}   
   \end{center}
   \caption{Quantum numbers of the fermion and scalar fields under the gauged subgroup.}
   \label{Tab:charges}
\end{table}

The effective potential for the SM Higgs generated by Eqs.~\eqref{eq:yukawa},~\eqref{eq:mass} was calculated in Ref.~\cite{Chacko:2005pe} and is finite.\footnote{Our normalization of the symmetry breaking scale $f$ differs from that of Ref.~\cite{Chacko:2005pe} by $\sqrt{2}$, i.e., $f_{\rm us} = \sqrt{2} f_{\rm CGH}$.} The vector-like mass $\tilde{M}$ plays the role of the cutoff to the logarithmically divergent contribution to the Higgs quartic term from the SM top and the twin top. Therefore it affects the physical Higgs boson mass. Of course, there could exist a ``bare'' $Z_2$ symmetric Higgs quartic term at the scale $\tilde{M}$ already,
\beq \label{Vkappa}
\kappa \left( |H_A|^4 + |H_B|^4 \right) \= \kappa \frac{f^4}{4} \left( \sin \left( \frac{h}{f} \right)^4 + \cos \left( \frac{h}{f} \right)^4 \right).
\eeq
In a UV-complete theory, it could arise from integrating out high energy physics above $\tilde{M}$, e.g., higher resonances or KK modes, or from a brane term in extra dimensional models.\footnote{If the underlying strong dynamics respects only $SU(4)$ global symmetry, the $\kappa$ term could arise at the order $\kappa \sim g_{\rm SM}^2$, where $g_{\rm SM}$ is an SM coupling representing the explicit breaking of $SU(4)$~\cite{Barbieri:2005ri,Barbieri:2015lqa}. On the other hand if the underlying strong dynamics respect an $SO(8)$ global symmetry which also protects the custodial $SU(2)$ symmetry, the $\kappa$ term will be suppressed by an additional loop factor ${g_{\rm SM}^2}/{(16\pi^2)}$~\cite{Chacko:2005un,Barbieri:2015lqa}.}

The complete Higgs potential can thus be written as $V = V_{\rm top}+ V_{\rm gauge} + V_{\rm \cancel{Z_2}} + V_{\kappa}$, where
$V_{\rm top}$ is the radiative Coleman-Weinberg (CW) contribution computed using Eqs.~\eqref{eq:yukawa},~\eqref{eq:mass}, $V_{\rm gauge}$ is the CW gauge contribution, whereas $V_{\rm \cancel{Z_2}}$ is given by Eq.~\eqref{softZ2break} and $V_\kappa$ by Eq.~\eqref{Vkappa}. With the extended top sector, $V_{\rm top}$ is finite and its one-loop CW contribution has been calculated in Ref.~\cite{Chacko:2005pe}. Because it depends on the fourth power of the top Yukawa coupling which has a strong scale dependence, the higher loop contributions are non-negligible and can significantly affect the Higgs mass prediction, analogous to the SUSY case.  The leading logarithmic corrections can be re-summed using renormalization group (RG) techniques. To include the leading higher loop contribution, 
we follow the results of Ref.~\cite{Haber:1996fp}, which demonstrated that in SUSY a good approximation to the RG-improved potential is obtained by replacing every occurrence of $m_t$ with the running top mass evaluated at the scale $\sqrt{m_t m_{\tilde{t}}}$, where $m_{\tilde{t}}$ is the stop mass. In our case, we shall instead use the running top mass evaluated at $\mu_t \equiv \sqrt{m_t \tilde{M}}$. Concretely, in $V_{\rm top}$ we replace $y_t$ with
\begin{equation}
y_t(\mu_t) = y_t(m_t) \left[1-\frac{1}{\pi}\left(\alpha_s - \frac{3}{16}\alpha_t\right)\log\frac{\mu_t^2}{m_t^2}\right]\,,
\end{equation}
which is defined by $y_t(\mu) = \sqrt{2} m_t(\mu) /v$ without the Higgs wave function renormalization.
The couplings on the right-hand side of the equation  are evaluated at $m_t$. 
In our calculation we use the following numerical values \cite{Buttazzo:2013uya}
\begin{equation}
m_t = 173.34\,, \qquad y_t(m_t) = 0.94018\,,\qquad g_s(m_t) = 1.1666\,,
\end{equation}
Notice that this procedure neglects corrections of order $y_t^2 f^2/(2\tilde{M}^2)$. 
The contribution to the Higgs potential from the gauge interactions $V_{\rm gauge}$ could also be made finite by extending the gauge sector. However, it is much smaller and has little dependence on the cutoff. We simply cut off its residual divergence at 10 TeV in our numerical calculation. The dependence on the twin hypercharge breaking scale is also mild and we set it to be $1\;\mathrm{TeV}$.
\begin{figure}[t]
 \begin{center}
\includegraphics[width=0.49\textwidth]{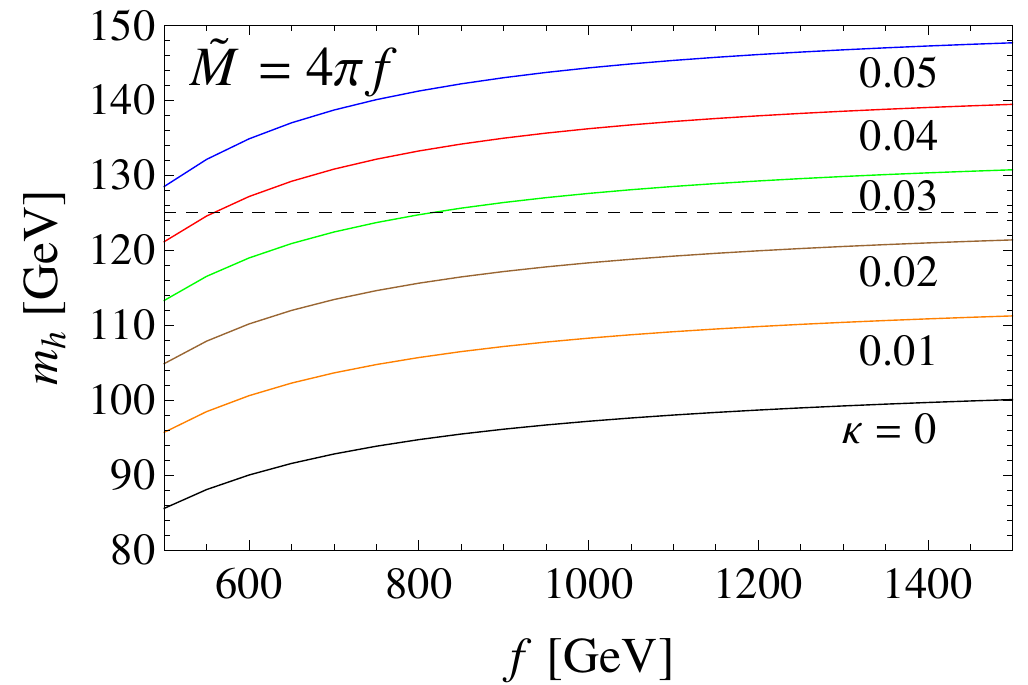}
\includegraphics[width=0.49\textwidth]{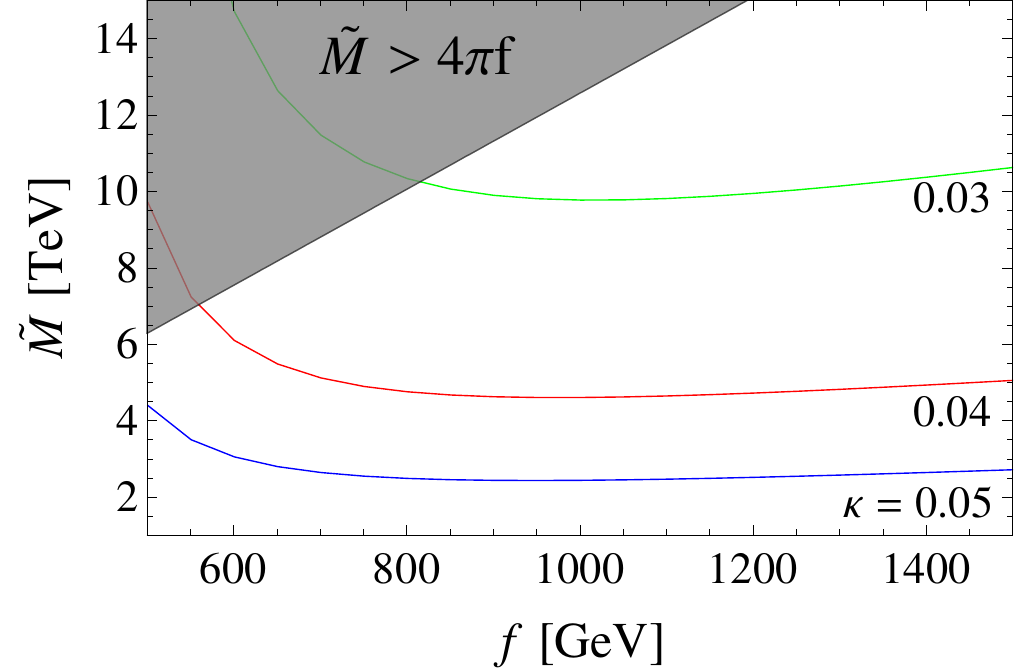}
 \end{center}
 \caption{Left panel: Higgs mass as a function of $f$, obtained setting $\tilde{M} = 4\pi f$. The dashed line indicates the observed value $m_h = 125\;\mathrm{GeV}$. Right panel: $\tilde{M}$ as a function of $f$, setting $m_h = 125\;\mathrm{GeV}$. In both panels the VEV is required to be at the physical value, $v\simeq 246\;\mathrm{GeV}$.}
\label{fig:Mkappa}
\end{figure}

The Higgs boson mass depends on the scales $\tilde{M}$ and $f$, which determine the loop contribution, as well as the bare quartic coupling $\kappa$ at scale $\tilde{M}$. To obtain insight on this interplay, in the left panel of Fig.~\ref{fig:Mkappa} we set $\tilde{M}=4\pi f$, the highest value that one can reasonably imagine, and show the Higgs mass as function of $f$ for a representative set of $\kappa$ values. We see that if the bare quartic is absent the Higgs is too light, $m_h \lesssim 100\;\mathrm{GeV}$, and $\kappa\gtrsim 0.03$ is required to reproduce the observed value $m_h = 125$ GeV. In the right panel of Fig.~\ref{fig:Mkappa} we show contours of $\kappa\geq 0.03$ in the $(f,\tilde{M})$ plane, after setting $m_h = 125\;\mathrm{GeV}$. We see that the exotic quark mass $\tilde{M}$ is quite sensitive to the bare quartic coupling $\kappa$. A slight increase in $\kappa$ above the minimum required value, $\tilde{M}$ can be reduced to be much lighter than the cutoff. Based on these considerations, in the rest of the paper we consider $\tilde{M}$ as a free parameter, only bounded from below by theoretical and experimental limits and from above by the cutoff  $4\pi f$.


\section{Exotic Quark Production and Decay}
\label{sec:production}
\begin{figure}
\centering
\begin{subfigure}{.50\textwidth}
  \centering
  \includegraphics[width=\linewidth]{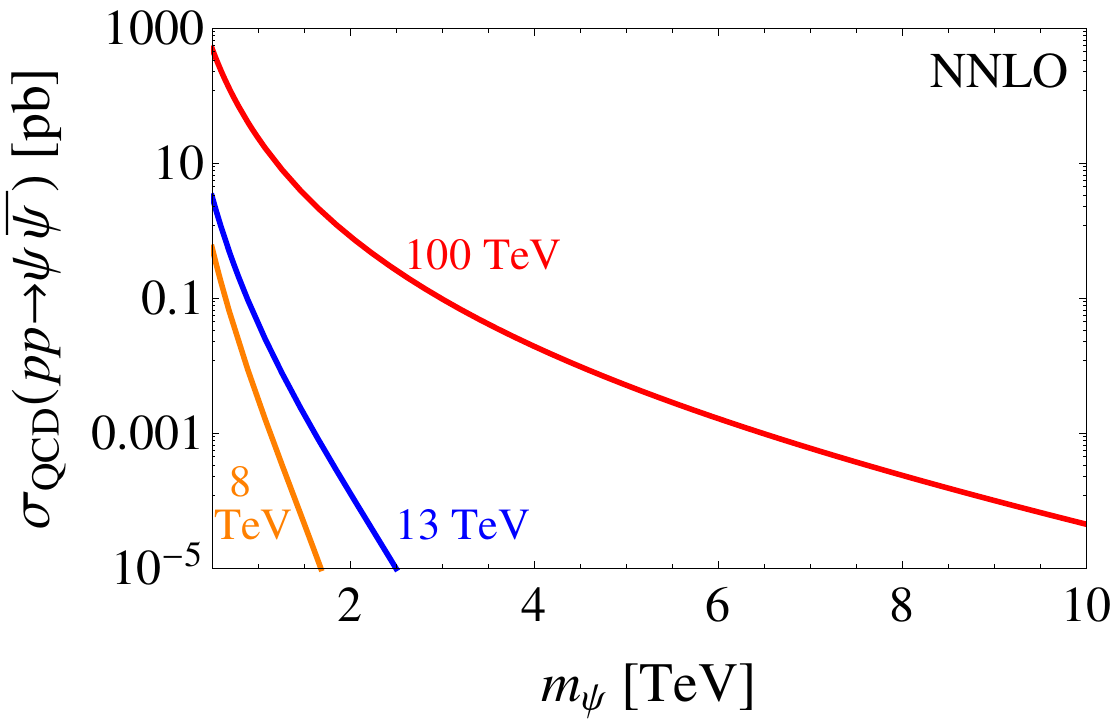}
  \label{fig:sub1}
\end{subfigure}\hspace{1mm}%
\begin{subfigure}{.48\textwidth}
  \centering
  \includegraphics[width=\linewidth]{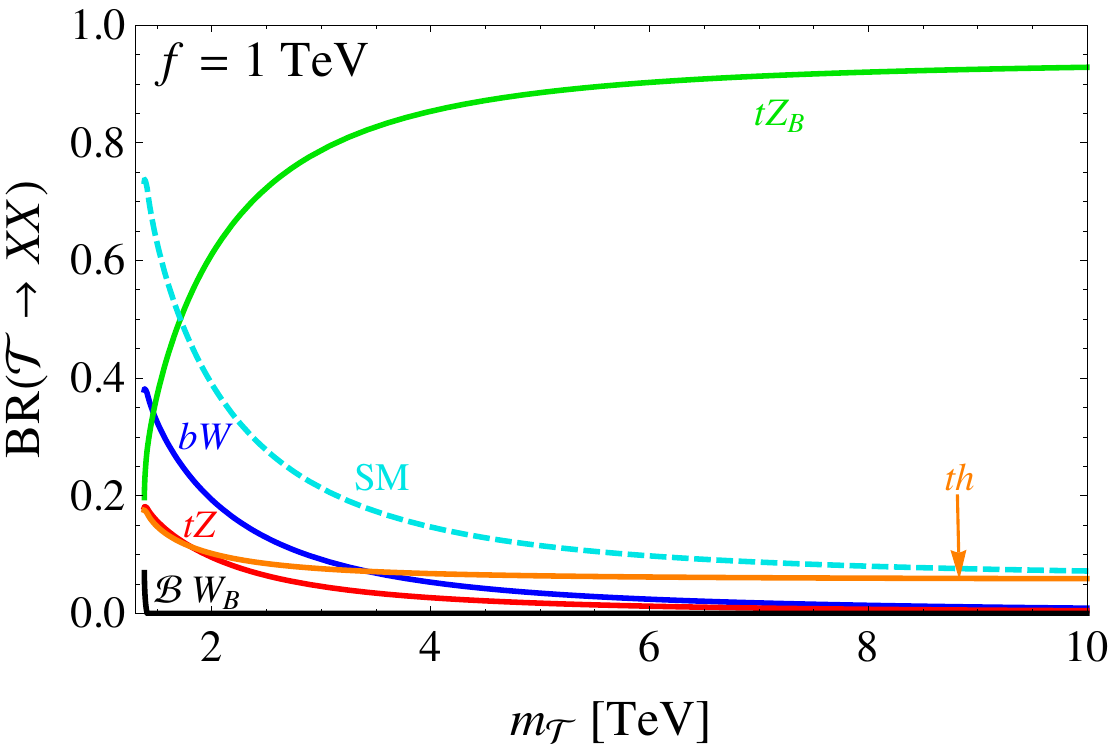}
  \label{fig:sub2}
\end{subfigure}
\caption{Left: cross section for QCD pair production of the exotic quark $\psi=\left\{\mathcal{T},\mathcal{B}\right\}$ at different collider energies. The numbers are computed at NNLO using Hathor2.0~\cite{Hathor}. Right: branching ratios of the exotic quark $\mathcal{T}$ for $f = 1\;\mathrm{TeV}$. The cyan dashed line indicates the sum of the $bW,tZ$ and $th$ branching ratios. The $\mathcal{B}$ instead only decays into $tW_B$.}
\label{fig:qcd_prod}
\end{figure}

Being colored, the exotic quarks $\tilde{u}^A_{3},\tilde{d}^A_{3}$ can be produced with large cross sections at hadron colliders, as shown in Fig.~\ref{fig:qcd_prod}. To study their phenomenology, we need to know how they decay. First we notice that $\tilde{q}^A_3$ and $u_{3R}^A$ transform in different representations of the twin electroweak group, but have identical SM quantum numbers. The exotic quark $\tilde{u}^A_3$ has zero twin electic charge, therefore it can mix with the top at $O(f)$. The relevant terms contained in $\mathcal{L}_t + \mathcal{L}_{m}$ are
\begin{equation} \label{eq:yukawas}
-\,y_t H_A^\dagger \overline{u}_{3R}^A q_{3L}^A -\,y_t H_B^\dagger \overline{u}_{3R}^A \tilde{q}_{3L}^A - \tilde{M} \overline{\tilde{q}}_{3R}^A \tilde{q}_{3L}^A + \mathrm{h.c.} \; \longrightarrow \; - \begin{pmatrix} \overline{u}_{3R}^A & \overline{\tilde{u}}_{3R}^{A} \end{pmatrix} \begin{pmatrix} \frac{y_t f}{\sqrt{2}}s_h & \frac{y_t f}{\sqrt{2}}c_h  \\ 0 & \tilde{M} \end{pmatrix} \begin{pmatrix} u^A_{3L} \\ \tilde{u}^{A}_{3L} \end{pmatrix} + \mathrm{h.c.}
\end{equation}
where we defined $s_h = \sin(h/f)$ and $c_h = \cos(h/f)$. The mass matrix, which we denote by $\mathcal{M}_f$, is diagonalized by the unitary transformations $U_{R,L}$, $U_R^\dagger \mathcal{M}_f U_L = \mathrm{diag}\;(m_t, m_{\mathcal{T}})$, with
\begin{equation}
\begin{pmatrix} u_{3R}^A \\ \tilde{u}^{A}_{3R} \end{pmatrix} = \begin{pmatrix} -c_R & s_R \\ s_R & c_R  \end{pmatrix} \begin{pmatrix} t_R \\ \mathcal{T}_R \end{pmatrix}\,, \qquad  \begin{pmatrix} u_{3L}^A \\ \tilde{u}^{A}_{3L} \end{pmatrix} = \begin{pmatrix} -c_L & s_L \\ s_L & c_L  \end{pmatrix} \begin{pmatrix} t_L \\ \mathcal{T}_L \end{pmatrix}\,,
\end{equation}
where the mass eigenstates $t$ and $\mathcal{T}$ are the observed top quark and the heavy exotic quark ($m_t < m_{\mathcal{T}}$). The masses are given approximately by 
\begin{equation}
m_t \simeq \frac{\tilde{M}}{\sqrt{\tilde{M}^2+ y_t^2 f^2/2}}\,\frac{y_t v}{\sqrt{2}}\,,\qquad m_{\mathcal{T}} \simeq \sqrt{\tilde{M}^2 + y_t^2 f^2/2}\, ,
\end{equation} 
where we only retained the leading order in $v$, and the mixing angles are given by
\begin{equation}
s_L = \frac{m_t}{\tilde{M}}s_R\,,\qquad s_R = \frac{y_t f/\sqrt{2}}{\sqrt{\tilde{M}^2 + y_t^2 f^2/2}} + O(v^2).
\end{equation}
For $\tilde{M}=0$ the mass matrix has a zero eigenvalue, therefore there is a minimum value of $\tilde{M}$ consistent with the observed value of the top mass   
\begin{equation}\label{eq:Mbound}
\tilde{M} \geq m_t \frac{f}{v} \sqrt{1+\sqrt{1-v^2/f^2}} = \sqrt{2}\, m_t \frac{f}{v} + O(v^2/f^2)\,.
\end{equation}
This is also a lower bound on the mass of $\tilde{d}_{3}^A \equiv \mathcal{B}$, which is simply given by $\tilde{M}$.
From Eq.~\eqref{eq:Mbound} one can derive a lower bound on the mass of $\mathcal{T}$,
\begin{equation} \label{eq:mqu_bound}
m_{\mathcal{T}}\geq \frac{m_t}{v}f\sqrt{2\left(1 + \sqrt{1-v^2/f^2}\,\right)-v^2/f^2} = 2\,\frac{m_t}{v}f + O(v^2/f^2).
\end{equation}
For $f=1\;\mathrm{TeV}$ the corresponding lower bounds are $\tilde{M}\gtrsim 0.99\;\mathrm{TeV}$ and $m_{\mathcal{T}}\gtrsim 1.38\;\mathrm{TeV}$. 

Using the Goldstone equivalence theorem, we can compute the decays of the exotic quarks from the top Yukawa coupling. Plugging into the first two terms of Eq.~\eqref{eq:yukawas} the expression of $H$ expanded up to $O(1/f)$, and subsequently performing the rotations on the fermions, we arrive to the leading terms controlling the decays of $\mathcal{T}$
\begin{align}
-y_t \Bigg[i \frac{c_R}{\sqrt{2}}\pi_7\, \overline{t}_R \,\mathcal{T}_{L} + \,&\, s_R \frac{-i\pi_5 + \pi_6}{\sqrt{2}}\overline{\mathcal{T}}_{R}\, \mathcal{B}_{L} + s_R \frac{-i\pi_1 + \pi_2}{\sqrt{2}}\overline{\mathcal{T}}_{R}\, b_{L} + i s_R \frac{\pi_3}{\sqrt{2}} \,\overline{\mathcal{T}}_{R}\, t_{L} \nonumber \\
-\,&\,\frac{c_R}{\sqrt{2}}\left(-\frac{v}{f} + s_L\right)\, h\, \overline{t}_R\, \mathcal{T}_{L} - \frac{s_R}{\sqrt{2}} h \,\overline{\mathcal{T}}_{R}\, t_L \Bigg]  + \mathrm{h.c.}\,.
\end{align} 
Thus the main decay widths of $\mathcal{T}$ are approximately given by
\begin{align}
\Gamma(\mathcal{T} \to b W_A) \,=\, 2\, \Gamma(\mathcal{T} \to t_A Z_A) =& \frac{m_{\mathcal{T}}}{32\pi} y_t^2 s_R^2 \nonumber \\
\Gamma(\mathcal{T} \to t h) \,=&\, \frac{m_{\mathcal{T}}}{32\pi} \left(\lambda_L^2 + \lambda_R^2 + \frac{4m_t}{m_{\mathcal{T}}}\lambda_L \lambda_R\right) \nonumber\\
\Gamma(\mathcal{T} \to t Z_B) \,=&\, \frac{m_{\mathcal{T}}}{64\pi} y_t^2 c_R^2 \nonumber
\end{align}
with
\begin{equation}
\lambda_L = \frac{y_t c_R}{\sqrt{2}}\left(- \frac{v }{f} + s_L\right)\,\qquad \lambda_R = \frac{y_t s_R}{\sqrt{2}}\,.
\end{equation}
These approximate formulas reproduce to good accuracy the full results computed in unitary gauge. In addition, if the decay $\mathcal{T} \to \mathcal{B} W_{B}$ is kinematically open, which only occurs for light $\mathcal{T},\, \mathcal{B}$, there is also a small width for it. In the limit $m_{\mathcal{T}}\gg y_t f$, mixing effects can be neglected and the decays are fixed by the interaction $-y_t H_B^\dagger \overline{u}_{3R}^A  \tilde{q}_{3L}^A$: $\mathcal{T}$ only decays into $t Z_B$ and into $t h$, and the latter has a coupling suppressed by $v/f$,
\begin{equation}
\Gamma(\mathcal{T} \to t Z_B) \simeq \frac{m_{\mathcal{T}}}{64\pi} y_t^2\,,\qquad \Gamma(\mathcal{T} \to t h) \simeq \frac{m_{\mathcal{T}}}{64\pi} y_t^2 \frac{v^2}{f^2}\,,
\end{equation} 
leading to $\mathrm{BR}(\mathcal{T} \to t Z_B)\simeq (1+v^2/f^2)^{-1} \sim 0.94$ for $f=1\;\mathrm{TeV}$. The branching ratios (computed using the full diagonalization of the masses and couplings) are shown in the right panel of Fig.~\ref{fig:qcd_prod}. Because the twin top partner $\hat{t}$ is heavier than $Z_B$ ($m_{\hat{t}}/m_{Z_B} = m_t/m_Z > 1$), the main decay modes of $Z_B$ are a pair of light twin fermions, $\hat{b},\hat{\tau}$ and $\hat{\nu}$. The branching ratio into $\bar{\hat{b}}\hat{b}$ is approximately equal to $60\%$ for $m_{\hat{b},\,\hat{\tau},\,\hat{\nu}}\ll m_{Z_B}$.   
 
The $\mathcal{B}$ instead decays with unity branching ratio into $t W_B$. (Despite its name, $\mathcal{B}$ has electric charge 2/3, same as the top quark.) The coupling is obtained again via the equivalence theorem,
\begin{equation}
y_t\, c_R \frac{-i \pi_5 + \pi_6}{\sqrt{2}} \overline{t}_R\, \mathcal{B}_{L}
\end{equation}
with partial width
\begin{equation}
\Gamma(\mathcal{B} \to t W_B)\simeq \frac{\tilde{M}}{32\pi} y_t^2 c_R^2\,.
\end{equation}
The only kinematically allowed decay of $W_B$ is into a  $\hat{\tau}$ and $\hat{\nu}$, via the twin tau Yukawa coupling
\begin{equation}
-\mathcal{L}_{\tau} = y_\tau \overline{\tau}_{3R}^A H_A^T \epsilon \ell_{3L}^A + y_\tau^B \overline{\tau}_{3R}^B H_B^T \epsilon \ell_{3L}^B
\end{equation}
where $\epsilon \equiv i\sigma^2$.
%

\subsection{Bounds from Fermionic Top Partner Searches}\label{sec:toppartner}
As shown in Fig.~\ref{fig:qcd_prod}, $\mathcal{T}$ has a non-negligible branching ratio into the SM final state $th$ as well as, for low masses, $bW, tZ$. These three final states are covered in searches for charge-$2/3$ fermionic top partners, which generically appear in composite Higgs and Little Higgs models. Taking for example the $8$ TeV CMS search for a pair-produced charge-$2/3$ top partner $T$ \cite{CMSTsearch}, the expected constraint on the $T$ mass is $\simeq 770\;\mathrm{GeV}$. The bound assumes that $T$ decays only into the final states $bW, tZ$ and $th$, but to a good approximation is independent of the specific partition of the branching ratios (see Table 5 in Ref.~\cite{CMSTsearch}). Recalling Eq.~\eqref{eq:mqu_bound}, for $f\gtrsim 750\;\mathrm{GeV}$ the lightest viable mass of $\mathcal{T}$ is $\gtrsim 1\;\mathrm{TeV}$. We thus conclude that there is no relevant constraint from the 8 TeV analysis. It is, however, interesting to estimate the projected bounds achievable with future data. For this purpose we perform a simple rescaling through the Collider Reach method \cite{ColliderReach}. We find that the current bound of $770\;\mathrm{GeV}$ translates into $m_T > 1.53\,(1.95)\;\mathrm{TeV}$ at the $13$ TeV LHC with $300\,(3000)\;\mathrm{fb}^{-1}$, and $m_T > 7.67\;\mathrm{TeV}$ at a $100\;\mathrm{TeV}$ collider with $1\;\mathrm{ab}^{-1}$. Taking into account the fact that in our model the rate into the relevant final states is reduced by $\mathrm{BR}(\mathcal{T} \to bW + tZ + th)^2$, we arrive to the projected reach (setting $f=1\;\mathrm{TeV}$)
\begin{align}
m_{\mathcal{T}} &\gtrsim 1.41\;\mathrm{TeV}\qquad\quad 13\;\mathrm{TeV},\, 300\;\mathrm{fb}^{-1}, \nonumber \\
m_{\mathcal{T}} &\gtrsim 1.70\;\mathrm{TeV}\qquad\quad 13\;\mathrm{TeV},\, 3000\;\mathrm{fb}^{-1},\qquad (\mathrm{fermionic\;top\; partner\; searches}) \nonumber \\
m_{\mathcal{T}} &\gtrsim 4.13\;\mathrm{TeV}\qquad\quad 100\;\mathrm{TeV},\, 1\;\mathrm{ab}^{-1}. 
\end{align}  
Considering the theoretical lower bound of Eq.~\eqref{eq:mqu_bound}, for $f=1\;\mathrm{TeV}$ the top partner searches at LHC Run 2 will not significantly constrain the parameter space of the model.

\subsection{Bounds from Stop Searches}\label{sec:stop}
Under the assumption that the decay products of $Z_B$ and $W_B$ escape the detector, effectively giving a missing transverse energy signature, the searches for stop direct production can set a bound on the mass of $\mathcal{T},\mathcal{B}$ via the process $pp\to \overline{\mathcal{T}}\mathcal{T},\overline{\mathcal{B}}\mathcal{B}\to \bar{t}t + \mathrm{MET}$. Similarly to fermionic top partner searches, while stop analyses at $8$ TeV do not place any relevant constraint on the model, their reach is expected to improve significantly with future data. Because $\mathcal{B}$ has mass $\tilde{M} < m_{\mathcal{T}}$ and $\mathrm{BR}(\mathcal{B}\to t W_B) > \mathrm{BR}(\mathcal{T}\to t Z_B) $, we first focus on $\mathcal{B}$ production. To estimate the future reach, we make use of the Snowmass study of Ref.~\cite{Snowmass}. From its Fig.~1-24 right and  assuming a neutralino mass equal to $m_{W_B}(f=1\;\mathrm{TeV})\simeq 316\;\mathrm{GeV}$, we extract the bounds $m_{\tilde{t}}> 1.17$ TeV at $14$ TeV with $300$ fb$^{-1}$ of data, and $m_{\tilde{t}}>6.04$ TeV at a $100$ TeV collider with $3$ ab$^{-1}$ of data. Within the Collider Reach approximation, this corresponds to $m_{\tilde{t}}> 1.11\,(1.49)\;\mathrm{TeV}$ at $13$ TeV with $300\,(3000)$ fb$^{-1}$, and $m_{\tilde{t}}>5.07$ TeV at $100$ TeV with $1$ ab$^{-1}$. Finally, by making use of the stop production cross sections at $13$ and $100$ TeV \cite{StopXsec}, we are able to set the bounds on $\tilde{M}$ by solving the following equations for $m$,\footnote{The ratio of the fermion to scalar QCD pair production cross section varies between $6$ and $8$ for the particle masses and collider energies considered here.}
\begin{align}
\sigma_{pp\to \tilde{t}\tilde{t}^\ast,\, 13\;\mathrm{TeV}}(m_{\tilde{t}} = 1.11\;\mathrm{TeV}) \,=&\, \sigma_{pp\to \overline{\mathcal{B}}\mathcal{B},\,13\;\mathrm{TeV}} (m), \nonumber \\
\sigma_{pp\to \tilde{t}\tilde{t}^\ast,\, 13\;\mathrm{TeV}}(m_{\tilde{t}} = 1.49\;\mathrm{TeV}) \,=&\, \sigma_{pp\to \overline{\mathcal{B}}\mathcal{B},\,13\;\mathrm{TeV}} (m), \nonumber \\
\sigma_{pp\to \tilde{t}\tilde{t}^\ast,\, 100\;\mathrm{TeV}}(m_{\tilde{t}} = 5.07\;\mathrm{TeV}) \,=&\, 2\,\sigma_{pp\to \overline{\mathcal{B}}\mathcal{B},\,100\;\mathrm{TeV}} (m).
\end{align}
In the $100$ TeV case we included a factor $2$ to approximately account for the contribution of $\mathcal{T}$, which for large masses is almost degenerate with $\mathcal{B}$ and has branching ratio into $t Z_B$ close to unity. On the contrary, in the $13$ TeV case the effect of $\mathcal{T}$ was neglected, because in the relevant mass range it is significantly heavier than $\mathcal{B}$, and has suppressed branching ratio into $tZ_B$. In this way we arrive at
\begin{align}
\tilde{M} &\gtrsim 1.43\;\mathrm{TeV}\;\;\;(m_{\mathcal{T}} \gtrsim 1.63\;\mathrm{TeV})\qquad\quad 13\;\mathrm{TeV},\, 300\;\mathrm{fb}^{-1}, \nonumber \\
\tilde{M} &\gtrsim 1.86\;\mathrm{TeV}\;\;\;(m_{\mathcal{T}} \gtrsim 2.00\;\mathrm{TeV})\qquad\quad 13\;\mathrm{TeV},\, 3000\;\mathrm{fb}^{-1},\qquad (\mathrm{stop\; searches}) \nonumber \\
\tilde{M} &\gtrsim 7.58\;\mathrm{TeV}\;\;\;(m_{\mathcal{T}} \gtrsim 7.62\;\mathrm{TeV})\qquad\quad 100\;\mathrm{TeV},\, 1\;\mathrm{ab}^{-1},\,
\end{align}  
where the bounds on $m_{\mathcal{T}}$ are obtained for $f=1\;\mathrm{TeV}$. Compared to top partner searches, stop searches will provide  comparable constraints at $13$ TeV and much stronger ones at a $100$ TeV collider, due to the decrease of $\mathrm{BR}(\mathcal{T}\to bW + tZ + th)$ at large masses. 

\section{Hidden Sector Phenomenology}\label{sec:hiddenpheno}
As we saw in the previous section, the dominant decay of the exotic quark $\mathcal{T}$ is into a SM top and a twin gauge boson $Z_B$. About 60\% of the $Z_B$'s then promptly decay into a pair of $\hat{b}$'s, which soon build up a twin QCD string and form a bound state. The bound state eventually undergoes twin hadronization and produces a number of twin hadrons, some of which can decay back to the SM with long lifetimes, giving rise to displaced vertices. Thus the signal we will be after is
\begin{equation}\label{exoquark signal}
pp\;\to\; (\mathcal{T}\to t Z_B)(\overline{\mathcal{T}}\to \bar{t}Z_B)\; \to\; t\bar{t} \;+\; \mathrm{twin\;hadrons}\,,\qquad \mathrm{twin\;hadron}\;\to\; \mathrm{displaced\;vertex}. 
\end{equation}
Triggering on these events is straightforward, for example by requiring one hard lepton from top decays, and the combination of the prompt $t\bar{t}$ pair and displaced vertex (DV) makes the signature essentially background-free. For $Z_B$ mass much larger than $m_{\hat{b}}$ and the twin confinement scale $\Lambda$, the initial $\hat{b}\bar{\hat{b}}$ bound state being produced is highly excited. From the standard color string picture with two $\hat{b}$'s connected by a tube of color flux, the potential is a linear function of the distance with
\begin{equation}
V(r)\simeq \sigma_s r
\end{equation}
with $\sigma_s \simeq 3\Lambda^2$ based on the lattice calculation~\cite{Bali:2000vr} (with a corresponding $\Lambda_{\rm QCD} \approx 250\, \mathrm{MeV}$). For $m_{Z_B}\simeq 360$ GeV and $\Lambda = 5$ GeV, the initial length of the string $r_{max}\sim m_{Z_B}/(3 \Lambda^2)$ is of order $10$ times longer than the size of lowest states ($\sim \Lambda^{-1}$). It is therefore necessary to include the dynamics of the string in estimating the final number of twin hadrons being produced. 

For concreteness, we consider the benchmark values for our study:
\begin{equation}\label{eq:benchmark}
f= 1\, \text{TeV}, \quad \Lambda =5\,\text{GeV},\quad m_{Z_B} = 360\,\text{GeV}\,.
\end{equation}
These values are motivated by naturalness and $f=1\;\mathrm{TeV}$ satisfies the current Higgs constraints, while having the Higgs mass tuning to be better than $10\%$. The Higgs tuning is insensitive to a large range of $m_{\hat{b}}$, and the signal for the exotic quark decay mainly depends on the relative sizes between $m_{\hat{b}}$ and $\Lambda$. 

The decay of the initial $\hat{b}\bar{\hat{b}}$ bound state either happens through string breakings or emissions of light twin hadrons. If $m_{\hat{b}}\ll \Lambda$, twin bottoms can be easily produced inside the string and then the excited bound state will be broken  into less energetic ones. In the opposite limit, $m_{\hat{b}}\gg \Lambda$, the production of twin quarks is exponentially suppressed, and the small acceleration $\Lambda^2/m_{\hat{b}}$ of the $\hat{b}$'s at the end of the string do not radiate enough energy to produce twin hadrons. The decay of the excited states then occur through emissions of twin glueballs or light twin hadrons from the scattering between the twin quarks. To obtain a more quantitative picture, here we use a simplified string model discussed in Ref.~\cite{Kang:2008ea} to study these two decay processes.

The time scale for the string breaking is determined by $m_{\hat{b}}/\Lambda$ with \cite{Kang:2008ea}
\begin{equation}\label{eq:stringbreak}
\tau_{\rm break}\sim\frac{4\pi^3}{K}e^{m_{\hat{b}}^2/\Lambda^2}\,.
\end{equation}
$K$ is the kinetic/potential energy of the string oscillation, which equals $m_{Z_B}$ for the initial hadron. Having a hard final state twin gluon emission in $Z_B\to\hat{b}\bar{\hat{b}}$ can change the energy, but the process is in the perturbative regime with a much suppressed probability. For the decay from hadron emissions, the time scale for the twin quark scattering $\hat{b}\bar{\hat{b}}\to\hat{b}\bar{\hat{b}}$ is given by
\begin{equation}
\tau_{\rm scatt}\sim \left(\frac{m_{Z_B}}{3\Lambda^2}\right)\,\left(\frac{\sigma_{\hat{b}\bar{\hat{b}}\to \hat{b}\bar{\hat{b}}+\text{hadron}}}{\Lambda^{-2}}\right)^{-1},\quad\sigma_{\hat{b}\bar{\hat{b}}\to\hat{b}\bar{\hat{b}}+\text{hadron}}\sim\frac{4\pi\hat{\alpha}_s^2}{m_{\hat{G}}^2}\,.  
\end{equation}
Here $m_{Z_B}/(3\Lambda^2)$ sets the time scale between each scattering for having two relativistic particles traveling through the string length $r_{max}\sim m_{Z_B}/(3\Lambda^2)$. The ratio $\sigma/\Lambda^{-2}$ gives the probability for the twin quarks to meet in the radial direction so the scattering can happen. The scattering cross section is estimated assuming the two $\hat{b}$'s exchange a $t$-channel twin gluon with energy of the order of the mass of the lightest twin hadron (taken here to be a twin glueball, $\hat{G}$). In this study, the emitted twin hadrons have masses much larger than $\Lambda$, so we assume that the perturbative description of the scattering process is applicable.\footnote{A complication which can spoil this simple picture is the change of the bound state angular momentum. Each twin hadron emission changes the size of momentum by $\sim m_{\hat{G}}\sqrt{\sigma_{\hat{b}\bar{\hat{b}}\to\hat{b}\bar{\hat{b}}+\text{hadron}}}\sim \sqrt{4\pi}\hat{\alpha}_s\lsim 1$. After a few scatterings, there can be an $O(1)$ variation of the orbital angular momentum. If the bound state has higher orbital angular momentum $\ell$, the hadron emission rate is suppressed by $\sim\ell^{-\ell-2}$ compared to the S-wave state, due to the wavefunction suppression. Approximating the $\ell$ change as a one-dimensional random walk gives on average a $\Delta\ell\lesssim 2$ change for $n_{\hat{B}}\leq 8$ emissions (this is the maximum number of glueballs that can be emitted for our benchmark parameters, see App.~\ref{app:multiplicity}), resulting in a $\gtrsim 6\%$ suppression of the glueball emission rate. Nevertheless, since as discussed below we focus on regimes where the separation between the timescales for string scattering and breaking is of two orders of magnitude, the possible delay of the emission process due to the angular momentum barrier does not change the picture.} 

The twin glueball emission dominates if $\tau_{\rm break} \gg \tau_{\rm scatt}$ and the string breaking dominates in the opposite limit. Given various uncertainties in our estimates, we base our discussion only in the two extreme limits -- the ``twin glueball case'' for $\tau_{\rm break} > 100\, \tau_{\rm scatt}$ where highly excited bound states mainly decay by emitting glueballs, and the ``twin bottomonium case'' for $\tau_{\rm scatt} > 100\, \tau_{\rm break}$ where the bound states decay by breaking the QCD-string into less excited states.   Later we find that the average energy of the first glueball emission is around 80 GeV, at which scale the twin QCD coupling is $\hat{\alpha}_{s}\simeq 0.2$. For these values the twin bottomonium case corresponds to $m_{\hat{b}}<8$ GeV in our benchmark, and the twin glueball case corresponds to $m_{\hat{b}}>17$ GeV. Many of the signals that we discuss later will apply to the range of $m_{\hat{b}}$ in between, but we are less certain which process will dominate.

Besides the twin QCD process, having a twin electroweak decay of the hadrons can greatly alter the phenomenology. This depends on the masses of the twin tau and twin neutrino, which are free parameters in the Fraternal Twin Higgs scenario. It has been argued that relatively heavy twin leptons can be a good dark matter candidate~\cite{Garcia:2015loa,Craig:2015xla}. If both the twin tau and twin neutrino are heavier than half of the hadron masses, various low lying bound states from the twin glueball or the twin bottomonium case can only decay into SM particles through the Higgs mediation or kinetic mixing. On the other hand, if the twin leptons are light so that the light twin hadrons can decay into them, then the collider signals depends on the relative branching fractions between the decays into twin leptons and into SM particles. In the case that decay into twin leptons dominates and the twin leptons are stable (on the collider time scale), they will escape the detector and leave missing energy as the only signal of the decay. However, stable light twin leptons generically have a relic density which over-closes the universe if they follow the standard thermal history~\cite{Garcia:2015loa,Craig:2015xla}, so it may be desirable to have them decay back to the SM particles through higher dimensional operators. If they have the appropriate decay length such that they decay inside the detector, there could also be exciting collider signals associated with twin leptons. The ``twin lepton case'' will be discussed in Sec.~\ref{sec:twintau}. 

\subsection{Twin Bottomonium Signals}\label{sec:bottom} 
If the twin quark mass is comparable to or smaller than the confinement scale $\Lambda$, the bound state formed right after the $Z_B$ decay will soon break into shorter twin color strings from the pair production of $\hat{b}\bar{\hat{b}}$ pairs in the string. To study the collider phenomenology we need to know how many twin bottomonia are produced at the end and their typical energy. In App.~\ref{app:multiplicity} we use a simplified string model to estimate the number of twin bottomonia produced, taking into account the fact that when a string breaks the energy is deposited in both the binding energy and the kinetic energy of the shorter strings. It is found that they are comparable and the number of produced mesons is close to half the number obtained by simply dividing the total energy by the twin bottomonium mass. Previous studies of hadronization in the context of Hidden Valley models \cite{Strassler:2006im} were performed in Refs.~\cite{Han:2007ae,Schwaller:2015gea} with similar results.

To identify the candidate displaced signals, we need to understand the properties of the twin bottomonia. In most of the following discussion we will assume that twin leptons are heavy, and play no role in the twin bottomonium decays. We will comment at the end on how light twin leptons would affect the exotic quark signals.  

Based on the SM $c\bar{c}$ and $b\bar{b}$ meson spectra, the lightest twin bottomonia are expected to be, in order of increasing mass, $\hat{\eta}_b$ ($0^{-+}$), $\hat{\Upsilon}$ ($1^{--}$) and $\hat{\chi}_{b0}$ ($0^{++}$).\footnote{This is true in the limit $m_{\hat{b}}\gg \Lambda$. In the opposite regime $m_{\hat{b}}\ll \Lambda$, according to lattice studies of $1$-flavor QCD \cite{Farchioni:2007dw} the $0^{-+}$ is lighter than the $0^{++}$. However, no results for the mass of the $1^{--}$ meson are currently available (we thank G.~M\"unster for clarifications about this point). For definiteness, in the following we assume $m_{\hat{\eta}_b}< m_{\hat{\Upsilon}} < m_{\hat{\chi}_{b0}}$ for all $m_{\hat{b}}$.} While the $\hat{\chi}_{b0}$ decays into SM particles through the Higgs portal, in the absence of other mediation channels the $\hat{\eta}_b$ and $\hat{\Upsilon}$ are meta-stable on cosmological time scales. Thus a universe where twin bottomonia are the lightest twin particles typically possesses a too large mass density. Requiring the decay of the lightest states to be sufficiently fast gives an interesting interplay between cosmological constraints and collider searches. Here we report the main results of our analysis, while several technical details are collected in App.~\ref{app:bottomonium}.

The proper decay length of the $\hat{\chi}_{b0}$ through the Higgs mixing is
\begin{eqnarray} \label{B0decay_quirk}
c\tau_{\hat{\chi}_{b0}}&\simeq& 8.3\,\text{cm}\left(\frac{m_b}{m_{\hat{b}}}\right)^5\left(\frac{f}{1\;\mathrm{TeV}}\right)^4\left(\frac{5\,\text{GeV}}{\Lambda}\right)^2\left[\frac{9}{5}\left(\frac{\sqrt{s}}{3m_{\hat{b}}}\right)^2-\frac{4}{5} \right]^{-1}\quad (m_{\hat{b}}\gg \Lambda),
\\ \label{B0decay_NDA}
c\tau_{\hat{\chi}_{b0}}&\simeq& 3.8\,\text{cm}\left(\frac{m_b}{m_{\hat{b}}}\right)^2\left(\frac{f}{1\;\mathrm{TeV}}\right)^4\left(\frac{5\,\text{GeV}}{\Lambda}\right)^5 \left(\frac{\sqrt{s}}{3\Lambda}\right)^{-2}\qquad\qquad\,\,\quad\,\, (m_{\hat{b}}\ll \Lambda),
\end{eqnarray}
valid in the range $2m_b < \sqrt{s}\ll m_h$, where $\sqrt{s}$ is the mass of the (excited) twin bottomonium (we neglected the small decay width into $\tau^+\tau^{-}$). 
Based on the SM quarkonium spectra, we estimate that the lightest twin bottomonium has a mass $\sim 2(m_{\hat{b}}+\Lambda)$. 
The value of $\sqrt{s}$ is therefore expected to range between $2(m_{\hat{b}}+\Lambda)$ and $4(m_{\hat{b}}+\Lambda)$, above which the string can split again into two light twin bottomonia. We will use $\sqrt{s}=3(m_{\hat{b}}+\Lambda)$ in Eqs.~(\ref{B0decay_quirk},~\ref{B0decay_NDA}) as the representative value for our estimate. The variation of the lifetime from scanning $\sqrt{s}$ within the allowed range is approximately a factor $2$ up or down, which can be taken as an uncertainty in our prediction. Notice that, differently from Ref.~\cite{Craig:2015pha}, where the $\hat{b}\bar{\hat{b}}$ bound state was treated as a non-relativistic quarkonium system, our discussion focuses on the case where the internal energy of the twin meson is comparable to the twin quark mass. Due to the geometrical suppression that follows from having a longer string, our $\hat{\chi}_{b0}$ lifetime is longer than the estimate in Ref.~\cite{Craig:2015pha} for the same value of $m_{\hat{b}}$. The decay is prompt when the temperature of the universe is $\sim m_{\hat{b}}$ ($\sim 10^{-8}$ sec) and releases the $\hat{\chi}_{b0}$ density into the SM sector. 

The vector bound state $\hat{\Upsilon}$ instead does not decay through the Higgs portal. It can decay, on the other hand, through the $U(1)_D$ gauge boson if it exists and has a kinetic mixing with the SM hypercharge gauge boson. Such a mixing term, $-(\epsilon/2) B_{\mu\nu}\hat{B}^{\mu\nu}$, will be induced at one loop by  
the exotic quarks $\tilde{q}^{A,B}_{3}$, with a typical size $\epsilon\sim g^{\prime\,2} N_c \log(\Lambda_{\rm UV }/\tilde{M})/(16\pi^2)\sim 10^{-3}$. This operator allows the $\hat{\Upsilon}$ to decay to SM fermions, via the mixing of the twin and SM photons. The corresponding decay length is
\begin{align}
c\tau_{\hat{\Upsilon}} \,&\simeq\, 1.5\,\text{cm}\,\left(\frac{m_b}{m_{\hat{b}}}\right)^3 \left(\frac{m_{\hat{A}}}{100\,\text{GeV}}\right)^4\left(\frac{10^{-3}}{\epsilon}\right)^2\left(\frac{5\,\text{GeV}}{\Lambda}\right)^2\left[\left(\frac{\sqrt{s}}{3m_{\hat{b}}}\right)^2+\frac{2}{9}\right]^{-1}\nonumber\\ &\qquad\qquad\qquad\qquad\qquad\qquad\qquad\qquad\qquad\qquad\qquad\qquad\qquad\quad\,(m_{\hat{b}}\gg \Lambda),\label{eq:1--decay_quirk}\\
c\tau_{\hat{\Upsilon}} \,&\simeq\, 1.3\,\text{cm}\, \left(\frac{m_{\hat{A}}}{100\,\text{GeV}}\right)^4\left(\frac{10^{-3}}{\epsilon}\right)^2\left(\frac{5\,\text{GeV}}{\Lambda}\right)^5\left(\frac{\sqrt{s}}{3\Lambda}\right)^{-2}\qquad\quad\;\; (m_{\hat{b}}\ll \Lambda),\label{eq:1--decay_NDA}
\end{align}
valid in the range $2m_b< \sqrt{s}\ll m_{\hat{A}}$ (twin photon mass). For an electroweak-scale twin photon, the current bound on $\epsilon$ mainly comes from electroweak precision measurements and the dilepton resonance searches \cite{Curtin:2014cca}, which require $\epsilon\lsim 10^{-2}$. For a very small $\epsilon$ and/or a very large $m_{\hat{A}}$, $\hat{\Upsilon}$ would decay outside the detector, leaving only missing energy signals at colliders.
However as we will see, the metastability of the pseudo-scalar $\hat{\eta}_b$ and the constraints from the big-bang nucleosynthesis (BBN) motivate that $\hat{\Upsilon}$ should decay inside the detector. 

The argument for the need of a relatively short $\hat{\Upsilon}$ decay length goes as follows: Since the pseudo-scalar $\hat{\eta}_b$ cannot decay through a single gauge boson,\footnote{The inner product between the derivative coupling of the pseudo-scalar and the kinetic mixing operator $\sim (g^{\mu\nu}p^2-p^{\mu}p^{\nu})$ vanishes \cite{Essig:2009nc}.} it decays into four SM fermions, via a one-loop triangle coupling to two off-shell twin photons. This process has a rate $\Gamma \sim \alpha^4 \epsilon^4 \Lambda^9 /(64\pi^3 m_{\hat{A}}^8)$, and the decay happens after the start of BBN ($\sim 0.1$ sec) when $\epsilon\lsim (m_{\hat{A}}/200\,\text{GeV})^2$. This decay dumps a large amount of entropy into the SM sector and can easily destroy the delicate BBN process. The number changing process $4\hat{\eta}_b\to 2\hat{\eta}_b$, which could reduce the matter density, freezes out too early to be effective if the mass of $\hat{\eta}_b$ is above $100$ keV \cite{Hochberg:2014dra}. Without introducing any other mediation processes, the only way to reduce the $\hat{\eta}_b$ density is to have them annihilate into a slightly heavier state, $\hat{\eta}_b\hat{\eta}_b\to\hat{\Upsilon}\hat{\Upsilon}$, followed by a fast decay of $\hat{\Upsilon}$ to the SM. Since the scattering rate from the twin strong interaction is much larger than Hubble at temperature $\sim m_{\hat{b}}$, the scattering remains effective when $T>\Delta m_{\hat{b}}$ between the two bound states. If the $\hat{\Upsilon}$ decay is prompt compared to Hubble, we can reduce the $\hat{\eta}_b$ density to an acceptable value. The comoving number of two mesons decreases with $\hat{Y}(T)\simeq\hat{Y}_0\,\exp[-\Gamma_{\hat{\Upsilon}}/H(T)]$, where the initial number $\hat{Y}_0$ is fixed by the SM-twin sector decoupling, which gives $\hat{Y}_0= n/s=0.27/g_{*S}(T\gsim m_{\hat{b}})\simeq 0.03$ \cite{Garcia:2015toa,Farina:2015uea}. This is a large number compared to $Y\sim 10^{-11}$ for a $10$ GeV DM particle that gives the observed relic density. To greatly reduce the $\hat{\eta}_b$ abundance, we need $\Gamma/H\gsim 1$ when the temperature is around the confinement scale $\Lambda$ of few GeV, which sets the decay lifetime
\begin{equation}
c\tau_{\hat{\Upsilon}}\lsim 10^{-9}\,\,\text{sec},\,\,\text{or}\;\;\lsim 30\,\,\text{cm}.
\end{equation}
Thus the cosmological constraints suggest that the decay of $\hat{\Upsilon}$ should happen inside the collider detectors.  

On the other hand, the $125$ GeV Higgs boson has a small branching fraction of decay into twin bottom quarks, which through twin hadronization can produce $\hat{\Upsilon}$'s. The $\hat{\Upsilon}$ displaced signals from Higgs production can be detected at colliders~\cite{Strassler:2006ri,Buckley:2014ika}. CMS has searched for generic long-lived particles ($X$) decaying to a lepton pair in the inner detector~\cite{CMS:2014hka}. The $X$ particles are assumed to be produced in pairs by the decay of a scalar resonance, which can also be the Higgs. The result is an upper bound on the production $\sigma(h)\times\mathrm{BR}(h\to XX)$ times the branching ratio BR$(X\to\ell^+\ell^-)$, as a function of the proper lifetime $c\tau_{X}$. Because $\hat{\Upsilon}$ decays through the kinetic mixing between the SM and twin photons, it has a sizable branching ratio into SM leptons. We can adopt this search result by identifying $X=\hat{\Upsilon}$. The relevant process is
\begin{equation}\label{hdecay1--}
pp\;\to\; h\; \to\;\hat{\Upsilon}\hat{\Upsilon},\qquad\hat{\Upsilon}\,\to\,(\mu^+\mu^-)_{\rm DV}\,,
\end{equation}
where DV indicates a displaced vertex. We focus on the dimuon final state because it provides the strongest constraint. (See the top left panel of Fig.~5 in Ref.~\cite{CMS:2014hka}.) 
Based on the string breaking model in App.~\ref{app:multiplicity}, the string breaking still dominates over the scattering process ($\tau_{\rm scatt} > 10\, \tau_{\rm break}$) in the $\hat{b}\bar{\hat{b}}$ system produced from the Higgs decay for $m_{\hat{b}}< 8\;\mathrm{GeV}$. In this range the average twin meson mass $3(m_{\hat{b}}+\Lambda)$ lies between $15$ and $39$ GeV.\footnote{Notice that, to remove the contribution of SM quarkonium resonances, CMS applied a $m_{\ell\ell}>15$ GeV cut, which is automatically satisfied for the range of twin meson masses we consider.}
For a very small $m_{\hat{b}}$, the Higgs decay may produce more than two twin hadrons, but in order to match to the search of Ref.~\cite{CMS:2014hka} we make the conservative assumption that only two bottomonia are produced in the twin hadronization. To obtain the final constraint we also need the fraction of the bottomonia produced being the vector $\hat{\Upsilon}$ states, $R_{\hat{\Upsilon}}$, which is a major uncertainty in our result.

To estimate $R_{\hat{\Upsilon}}$, we again resort to the string breaking model.
The original $\hat{b}\bar{\hat{b}}$ pair produced from the Higgs decay does not carry any angular momentum. However, when the string is broken by the creation of the $\hat{b}\bar{\hat{b}}$ pair, that can happen at a point away from the center of the string cross section and the newly produced twin quark could pick up an impact factor $\sim\Lambda^{-1}$ with respect to the quark at the string end. The two twin bottom quarks of the final twin meson typically have a relative kinetic energy $\sim m_{\hat{b}}+\Lambda$, so the orbital angular momentum is expected to be ${\cal O}(1)$. States with higher orbital angular momenta also have larger masses and their production is suppressed. If we assume that all states with orbital angular momenta up to some maximum value $\ell$ are equally produced, then there will be a fraction $3/[4(\ell+1)^2]$ of them being the $\hat{\Upsilon}(1^{--})$ mesons.\footnote{For $\ell \geq 1$ there are other vector states which may also decay back to SM particles. Their decays are suppressed by the nonzero orbital angular momentum and for a conservative estimate we do not include them.} For $\ell = 0,\, 1,\, 2$ we obtain $R_{\hat{\Upsilon}}= 3/4, \, 3/16,\, 3/36$. In our study we take $R_{\hat{\Upsilon}}=3/16$ as the central value but also show the results for the other two values as the optimistic and pessimistic bounds. 

\begin{figure}
\begin{center}
\includegraphics[width=8cm]{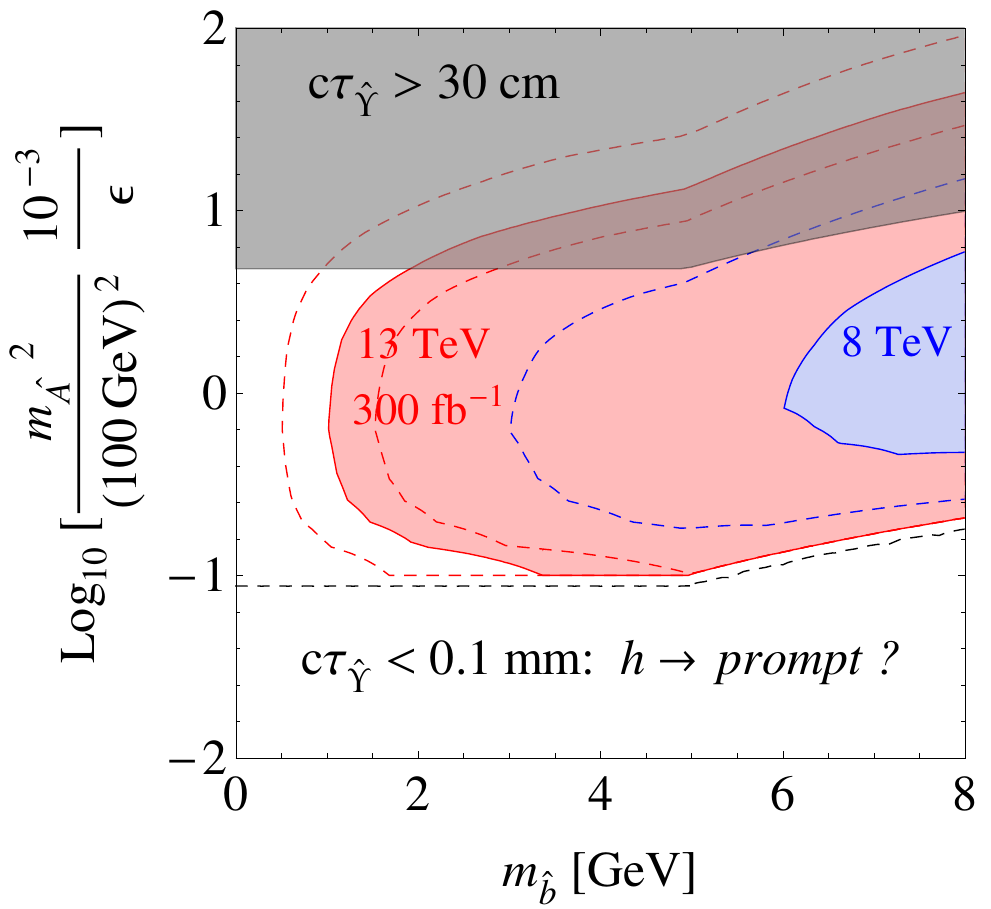}
\end{center}
\caption{Bounds on the twin photon mediation of the $\hat{\Upsilon}$ decay from the $h\to\hat{\Upsilon}\hat{\Upsilon}\to (\mu\mu)_{\mathrm{DV}}$ + anything process, at $8$ TeV (blue) and $13$ TeV (red). The solid contours assume that the probability to form two $\hat{\Upsilon}$ in the hadronization following the $h\to \hat{b}\bar{\hat{b}}$ decay is $3/16$, while the dashed contours correspond to $3/36$ and $3/4$. Also shown are the region where the lifetime of $\hat{\Upsilon}$ is too long to efficiently deplete the $\hat{\eta}_b$ density, which overcloses the universe (grey), and the contour corresponding to a $\hat{\Upsilon}$ lifetime of $0.1$ mm, where the displaced search loses sensitivity (dashed black). The kink at $m_{\hat{b}}=\Lambda =5$ GeV corresponds to the transition between the two expressions of $c\tau_{\hat{\Upsilon}}$ in Eqs.~(\ref{eq:1--decay_quirk},~\ref{eq:1--decay_NDA}).}\label{fig:highsto1--}
\end{figure}

For a fixed $\Lambda$, the decay length in Eqs.~(\ref{eq:1--decay_quirk},~\ref{eq:1--decay_NDA}) is mainly determined by the suppression scale $m_{\hat{A}}^2/\epsilon$, with some additional dependence on $m_{\hat{b}}$ for $m_{\hat{b}}\gg \Lambda\,$. On the other hand, the branching ratio for $h\to \bar{\hat{b}}\hat{b}$ is sensitive to $m_{\hat{b}}$. The region of the $(m_{\hat{b}},m_{\hat{A}}^2/\epsilon)$ plane excluded by the $8$ TeV search (assuming $R_{\hat{\Upsilon}}=3/16$) is shaded in blue in Fig.~\ref{fig:highsto1--}, where we also show (pink shaded region) the estimated exclusion with $300$ fb$^{-1}$ of $13$ TeV data. To obtain this projection we simply rescaled the $8$ TeV bound on the signal production rate by the luminosity ratio, which is a reasonable approximation for a background-free search. Interestingly, the LHC bound is complementary to the one coming from the requirement of depleting the $\hat{\eta}_b$ density: the cosmological bound disfavors a long decay length (grey shaded area in Fig.~\ref{fig:highsto1--}), whereas the $13$ TeV search can exclude most of the remaining region where the $\hat{\eta}_b$ is long-lived on collider scales. For smaller $m_{\hat{A}}/\sqrt{\epsilon}$ the $\hat{\Upsilon}$ decay length becomes shorter than $0.1$ mm, and the displaced dilepton search loses sensitivity. In this region additional constraints may come from, for example, the (prompt) $h\to 4\ell$ measurement. 

Having discussed the properties of the lowest-lying twin bottomonia, we return to our analysis of the exotic quark signals in Eq.~\eqref{exoquark signal}. We will study the possible displaced signatures arising from $\hat{\Upsilon}$ and $\hat{\chi}_{b0}$. The signals that we consider are 
\begin{equation}\label{exoquark signal_mesons}
pp\,\to\, (\mathcal{T}\to t Z_B)(\overline{\mathcal{T}}\to \bar{t}Z_B)\, \to\; t\bar{t} \,+\, \mathrm{twin\;bottomonia}\,,\; \hat{\Upsilon}\,\to\,(\mu^+\mu^-)_{\rm DV}\mbox{ or } \hat{\chi}_{b0}\;\to\; (b\bar{b})_{\mathrm{DV}}. 
\end{equation}
As discussed at the beginning of the section, the range of masses where twin bottomonium production dominates is $m_{\hat{b}}<8$ GeV. The number of twin bottomonia produced from the $Z_B$ decay, $n_{\hat{B}}$, is assumed to be $M_{Z_B}/2$ divided by the typical twin bottomonium mass $3(m_{\hat{b}}+\Lambda)$ and rounded down to an even number. (Having an even number will be more convenient in dealing with the kinematics of the $Z_B$ decay, as discussed later.) For $m_{\hat{b}}\in (0,8)$ GeV we find $n_{\hat{B}}\in (10,4)$. For $\hat{\Upsilon}$, the decay length and hence the search reach also depends on the parameter $m_{\hat{A}}^2/\epsilon$. We will assume $m_{\hat{A}}^2/\epsilon= (100\, \mathrm{GeV})^2/ 10^{-3}$ motivated by the cosmological constraint. The reach will be degraded if the decay length becomes significantly longer or shorter. On the other hand, $\hat{\chi}_{b0}$ decays through the Higgs portal which does not depend on the extra parameter and hence the result is less model-dependent.
In the following we estimate the reach of the exotic quark mass that can be obtained by searching for the process in Eq.~\eqref{exoquark signal_mesons} at the LHC and at a future $100$ TeV collider.

The production rate and decay branching ratios of the exotic quarks have been computed previously, and were shown in Fig.~\ref{fig:qcd_prod}. To calculate the number of expected signal events we need to estimate the probability of producing a $\hat{\Upsilon}$ or $\hat{\chi}_{b0}$ in the $Z_B\to \hat{b}\bar{\hat{b}}$ decay, as well as the efficiency for the DV arising from the twin bottomonium to pass the experimental cuts. For this purpose, we combine a simulation of the process with a simplified analytical approximation of the twin hadronization that follows the formation of the $\hat{b}\bar{\hat{b}}$ bound state. We treat the $\hat{\Upsilon}$ and $\hat{\chi}_{b0}$ signals separately, therefore we discuss the former first and comment later on the differences between the two analyses. 

We generate a parton-level sample of $pp\to (\mathcal{T}\to t Z_B)(\overline{\mathcal{T}}\to \bar{t}Z_B)$ with both $Z_B\to \hat{b}\bar{\hat{b}}$, using MadGraph5 \cite{MG5} and employing a model created with FeynRules 2.0 \cite{Alloul:2013bka}. Since we expect the twin bottomonium emission to happen with larger probability along the string direction, we boost each $\hat{b}\bar{\hat{b}}$ pair to the mother $Z_B$ rest frame, and assume that $n_{\hat{B}}/2$ produced bottomonia move in the $\hat{b}$ direction, and the other $n_{\hat{B}}/2$ move in the $\bar{\hat{b}}$ direction. To simplify the estimation, we further assume that the $Z_B$ energy is equally split among the twin bottomonia, a fraction $R_{\hat{\Upsilon}}$ of which is assumed to be $\hat{\Upsilon}$. As discussed above, we take $R_{\hat{\Upsilon}}=3/16$ to be our central estimate, and $3/4,3/36$ as the optimistic and pessimistic values, respectively. Once we boost back to the laboratory frame, the four momentum of each $\hat{\Upsilon}$ together with the proper lifetime in Eqs.~(\ref{eq:1--decay_quirk},~\ref{eq:1--decay_NDA}) determines the probability distribution for its decay, which is integrated over the volume of the detector in which the DV can be reconstructed. We focus on the cleanest decay $\hat{\Upsilon}\to \mu^+ \mu^-$ and thus take as reference the CMS search of Ref.~\cite{CMS:2014hka}, where dimuon DVs were searched for only in the inner detector (ID). This is sufficient to retain most of the signal, because for our benchmark parameters, the lifetime of $\hat{\Upsilon}$ is $O({\rm cm})$ with mild dependence on $m_{\hat{b}}$, see Eqs.~(\ref{eq:1--decay_quirk},~\ref{eq:1--decay_NDA}). We model the ID as an annulus with radii $1< r <50$ cm and efficiency for $(\mu^+ \mu^-)_{\rm DV}$ identification equal to a constant $50\%$. The cuts $|\eta|< 2.5$ and $p_T > 20$ GeV are also imposed on each $\hat{\Upsilon}$, to approximately reproduce the real experimental requirements on the muons. With the above procedure, we determine the total efficiency for observing a DV. On the other hand, the $t\bar{t}$ pair is assumed to decay semileptonically, and we include a $90\%$ efficiency for the lepton to pass basic selection cuts. Notice that while our simulation was performed using the process $(\mathcal{T}\to t Z_B)(\overline{\mathcal{T}}\to \bar{t}Z_B)$ with both $Z_B$ decaying into twin bottoms, in the rate calculation we also included all the other decay patterns that give rise to one hard lepton plus twin bottomonia, taking into account the appropriate rescaling factors. For example, we included the case where one of the exotic quarks decays to $bW,tZ$ or $th$, whose branching fractions are significant for low $m_{\mathcal{T}}$ and thus for estimating the LHC reach. 

The analysis of the $\hat{\chi}_{b0}$ displaced decays is similar, albeit with a few notable differences. For the $\hat{\chi}_{b0}$ fraction of produced bottomonia $R_{\hat{\chi}_{b0}}$, 
we take $R_{\hat{\chi}_{b0}}=1/16 \, (\ell=1),\,1/36\, (\ell=2)$ as the optimistic and pessimistic assumptions, respectively. (If only $\ell=0$ is available, the $P$-wave $\hat{\chi}_{b0}$ cannot be produced.) The $\hat{\chi}_{b0}$ decays dominantly into $b\bar{b}$, with lifetime that depends rather strongly on the mass of the twin bottom, see Eqs.~(\ref{B0decay_quirk},~\ref{B0decay_NDA}), and in particular becomes of $O({\rm meter})$ for $m_{\hat{b}}\lesssim 1\;\mathrm{GeV}$. Therefore it is important to include the reconstruction of $(b\bar{b})_{\rm DV}$ at radii larger than the size of the inner detector. For this purpose, we make reference to the ATLAS analyses of Refs.~\cite{Aad:2015uaa,Aad:2015asa}, where hadronic DVs were searched for in the ID, hadronic calorimeter (HCAL) and muon spectrometer (MS). The detector is thus roughly modeled as the sum of two annuli with radii $1< r <28$ cm and $200 < r < 750$ cm, representing the ID and HCAL+MS, respectively, with an efficiency for DV identification equal to a constant $10\%$.\footnote{In Ref.~\cite{Aad:2015uaa} it was found that the efficiency for hadronic DV reconstruction in the ID becomes very suppressed for $r < 1$-$4$ cm, depending on the signal model (see also Table~\ref{tab:dijetDV-param} later). Since our signal also contains a prompt $t\bar{t}$ pair that can be exploited to further suppress the background, we assume that reconstruction with $O(10)\%$ efficiency can be achieved down to $r = 1$ cm, which matches the minimum value for the dilepton DV search.} The cuts $|\eta|< 2.5$ and $p_T > 20$ GeV are also imposed on each $\hat{\chi}_{b0}$.

\begin{figure}
\begin{center}
\vspace{-2mm}
\includegraphics[width=11cm]{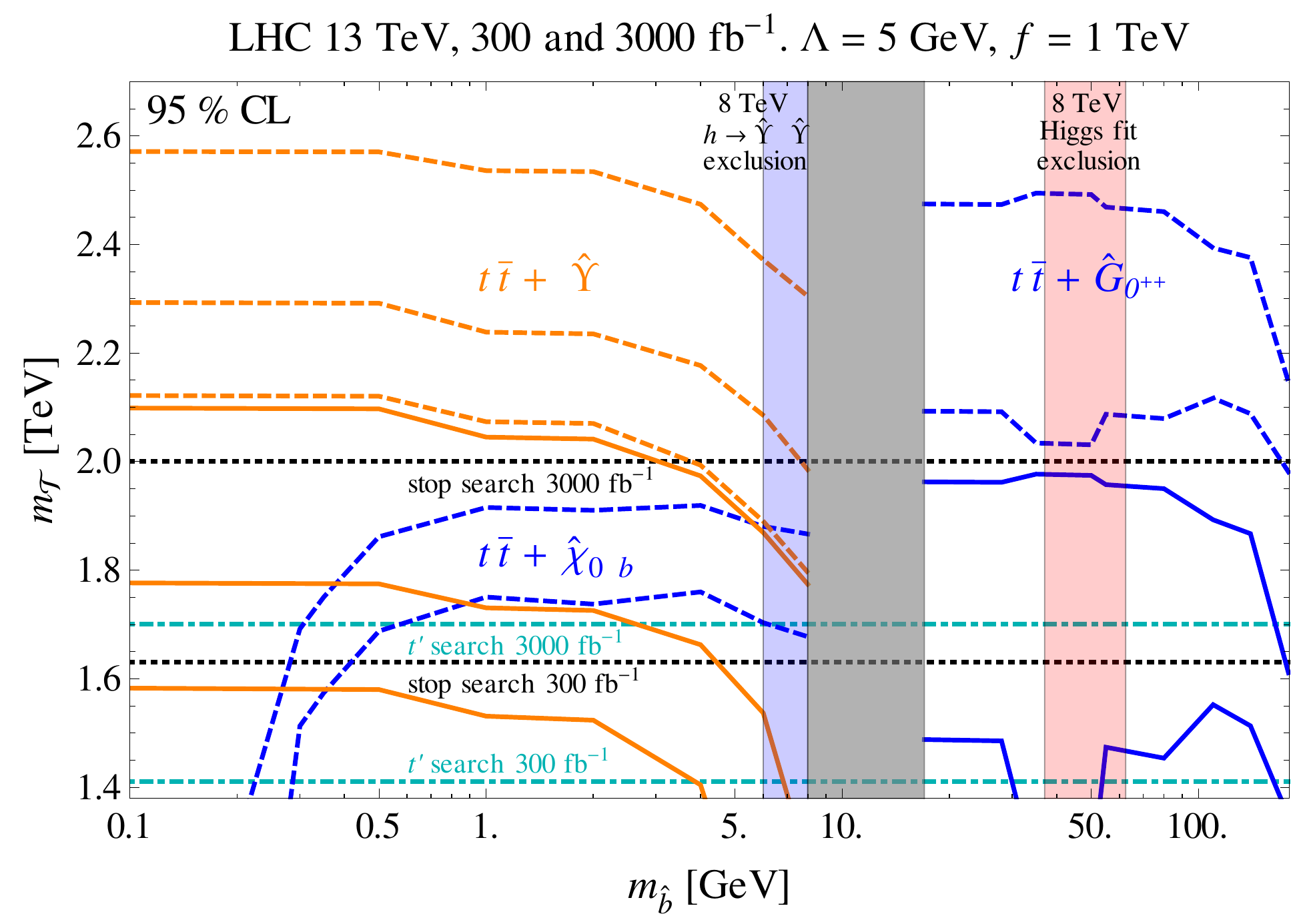}\\
\includegraphics[width=11cm]{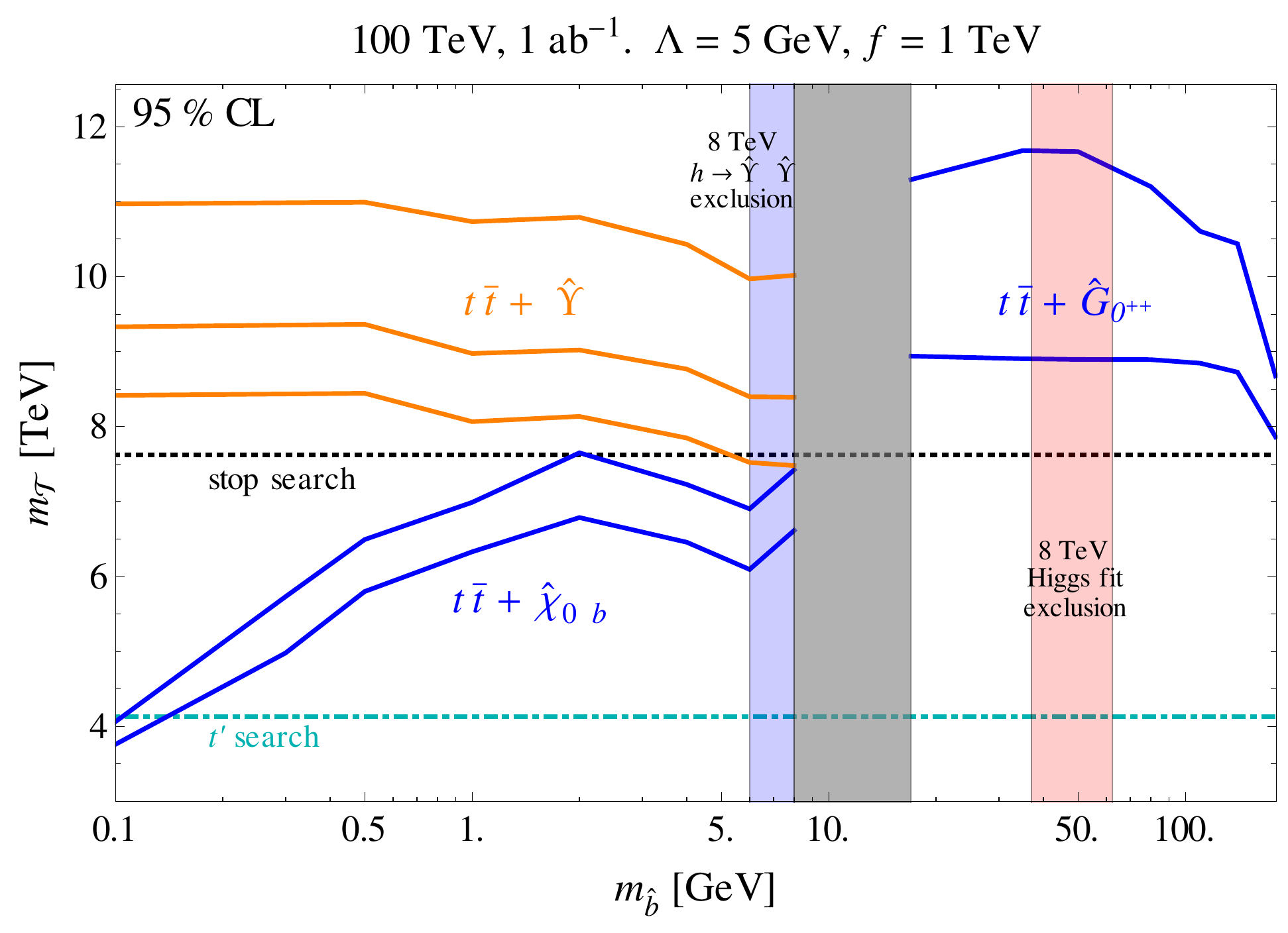}
\caption{Projected bounds on the mass of the exotic quark $\mathcal{T}$ at the 13 TeV LHC (top) and $100$ TeV collider (bottom), as functions of the mass of the twin bottom $\hat{b}$. The orange and blue curves correspond to the $t\bar{t}+$displaced twin hadron signals. In the top panel, solid~(dashed) curves correspond to a luminosity of $300\,(3000)$ fb$^{-1}$. The different curves for each signal illustrate the uncertainty in the estimate of the twin hadron fractions. Twin bottomonium (glueball) production dominates for light (heavy) $\hat{b}$, whereas in the gray shaded region both types of twin hadrons appear. See Sec.~\ref{sec:bottom} and \ref{sec:glueball} for more details. Also shown are the bounds from top partner (dot-dashed light blue lines) and stop (dotted black lines) searches, described in Sec.~\ref{sec:toppartner} and \ref{sec:stop}, respectively. The region shaded in blue (red) is already excluded by the $8$ TeV searches for Higgs displaced decays (global Higgs fit).
}\label{fig:glueball} 
\end{center}
\end{figure}
%

As already mentioned, both for the $t\bar{t} + \hat{\Upsilon}$ and $t\bar{t} +\hat{\chi}_{b0}$ signals we expect the SM background to be negligible. Therefore the $95\%$ CL lower bound on $m_{\mathcal{T}}$ is obtained by simply requiring the number of signal events to equal $3$. The result is shown in Fig.~\ref{fig:glueball} as a function of the $\hat{b}$ mass, in the upper panel for the 13 TeV LHC with $300$ fb$^{-1}$ (solid) and $3000$ fb$^{-1}$ (dashed), and in the lower panel for a $100$ TeV collider with 1 ab$^{-1}$. The reach of the $\hat{\Upsilon}$ search is shown in orange and the one of the $\hat{\chi}_{b0}$ search is shown in blue, with each curve corresponding to a different value of $R_{\hat{\Upsilon}}$ or $R_{\hat{\chi}_{b0}}$. In the same figure we also show the projected bounds from stop (black dotted lines) and fermionic top partner searches (dot-dashed light blue lines), as well as the region excluded by the $h\to\hat{\Upsilon}\hat{\Upsilon}$ search at $8$ TeV, $6 < m_{\hat{b}}< 8$ GeV (shaded in blue) from Fig.~\ref{fig:highsto1--}. Based on our assumptions, the $\hat{\Upsilon}$ signal could provide a better reach on the exotic quark mass compared to the stop searches. If we take the optimistic value of the $\hat{\Upsilon}$ fraction of produced bottomonia, $m_{\mathcal{T}}\sim 2.1\,(2.6)$ TeV can be excluded at the $13$ TeV LHC with $300\,(3000)$ fb$^{-1}$, while at a future $100$ TeV collider the sensitivity reaches $\sim 11$~TeV with $1$ ab$^{-1}$, essentially covering the full range up to $4\pi f$. Note that the expected reach from the twin photon-mediated $\hat{\Upsilon}$ decay depends on the ratio of the twin photon mass-squared and the kinetic mixing parameter. The numbers obtained here are based on $m_{\hat{A}}^2/\epsilon = (100\;\mathrm{GeV})^2/10^{-3}$. On the other hand, the $\hat{\chi}_{b0}$ search can be sensitive to $m_{\mathcal{T}}\sim 1.9$ TeV at the high-luminosity LHC, and up to $\sim 7$ TeV at the $100$ TeV collider. Estimates for luminosities different from 1 ab$^{-1}$ at the $100$ TeV collider can be easily obtained using the Collider Reach tool \cite{ColliderReach}, which assumes the signal rate scales with the partonic luminosities. This is a reasonable first approximation, because at large $m_{\mathcal{T}}$ the exotic quark branching ratios are approximately independent of the mass (see Fig.~\ref{fig:qcd_prod}), and the variation of the typical twin hadron boost factor only gives subleading corrections.      

So far we have assumed that the twin leptons are heavy and hence irrelevant in the twin bottomonium decay. If the twin leptons are light, e.g., lighter than half of the mass difference between $\hat{\chi}_{b0}$ and $\hat{\Upsilon}$, the $\hat{\chi}_{b0}$ can also decay through an off-shell $Z_B$ into $\hat{\Upsilon}\hat{\ell}\hat{\ell}$, where $\hat{\ell}$ denotes a twin lepton. The $\hat{\Upsilon}$ decays into $\hat{\ell}\hat{\ell}$, again through off-shell $Z_B$ exchange. Therefore the DV signals discussed above are expected to be suppressed. If the twin leptons are stable on collider scales, the $Z_B$ decay produces only missing energy. On the other hand, if through higher dimensional operators they can have fast enough decays back to the SM sector, their signals may also be detected. We will discuss the twin lepton decay in Sec.~\ref{sec:twintau}.        

\subsection{Twin Glueball Signals}\label{sec:glueball}
In the $m_{\hat{b}}\gg\Lambda$ limit the twin QCD string oscillates without breaking, and the energy of the bound state is reduced via emissions of twin glueballs. Since the glueball masses are much larger than the confinement scale $\Lambda$, we analyze the twin gluon emission that leads to twin glueball production using perturbation theory. To estimate the number of glueballs produced, we need to compute the average energy of a twin gluon emission at each scattering. We describe the process in the limit where the emitted twin gluon is soft and collinear to the string direction: when a pair of $\hat{b}$'s carrying a kinetic energy $E_0$ scatters, the probability of emitting a twin gluon with energy $E$ between $E_1<E<E_0$ can be approximated by
\begin{equation}\label{eq:radprob}
P(r)\simeq1-\exp\left[-\frac{\alpha_d}{\pi}\,(\ln\,r^{-2})^2\right],\quad r\equiv\frac{E_1}{E_0}.
\end{equation}
We assume that the hadronization process eventually turns the emitted twin gluon into a twin glueball if it has sufficient energy. With these assumptions, we estimate the average of the maximum number of twin glueballs that can be produced in the string de-excitation, which consists of several scatterings, until the internal energy of the string is lower than the lightest twin glueball mass.\footnote{The ``average'' in the sentence comes from the fact that we can only derive the probability of emitting twin gluons harder than a given energy $E_1$. By taking the average of $E_1$, we can estimate the lower bound on the energy emission in each scattering, and thus calculate the maximum number of twin glueballs produced.} In addition, if $m_{\hat{b}}$ is sufficiently large, after the last twin glueball emission the remnant string can annihilate into two twin glueballs. The details of the computation are provided in App.~\ref{app:multiplicity}, where we find that for our benchmark $(\Lambda,\,m_{Z_B})=(5,\,360)$ GeV the de-excitation of the $\hat{b}\bar{\hat{b}}$ string produces a total of $n_{\hat{G}}\in (8,2)$ twin glueballs in the relevant range $m_{\hat{b}}\in (17,m_{Z_B}/2)$. We then turn to a description of the properties of the twin glueballs, which is needed to identify their possible displaced signatures. We first assume that the twin leptons are heavy, and comment on how having light leptons would affect the exotic quark signals at the end. 

Based on lattice computations in pure-glue QCD \cite{Morningstar:1999rf,Chen:2005mg}, the three lightest glueballs are known to be, in order of increasing mass, $\hat{G}_{0^{++}}$, $\hat{G}_{2^{++}}$ and $\hat{G}_{0^{-+}}$, all composed of two gluons. Their masses are related by $m_{0^{-+}}=1.5\, m_{0^{++}}$ and $m_{2^{++}}=1.4\, m_{0^{++}}$, where $m_{0^{++}} \simeq 6.8\Lambda$ is the mass of the $0^{++}$ glueball. The decay constants of the glueballs, denoted by $f_i$ $(i=0^{++},\,\,2^{++},\,\,0^{-+})$, can be expressed in terms of their masses as
\begin{align}\label{eq:glueballmass}
4\pi \hat{\alpha}_s f_{i} = C_i\,m_{i}^3\,,
\end{align} 
with 
$C_i=(3.1,\,0.038,\,0.51)$. The gluonic operators corresponding to the decay constants are given in Eq.~(1) of Ref.~\cite{Chen:2005mg}.

The $\hat{G}_{0^{++}}$ decays into SM particles through its mixing with the Higgs, which is mediated by the twin top loop~\cite{Craig:2015pha}. In the limit $m_h \ll 2 m_{\hat{t}}$, the effective coupling is
\begin{equation}
\mathcal{L}_{\mathrm{eff}}= -\frac{\hat{\alpha}_s}{12\pi}\frac{v}{f}\frac{h}{f}\hat{G}_{\mu\nu}^a \hat{G}^{\mu\nu\,a}.
\end{equation} 
This leads to a width for the decay $\hat{G}_{0^{++}}\to YY$ ($Y$ denotes a SM particle)  
\begin{equation} \label{glueballdecay}
\Gamma(\hat{G}_{0^{++}}\to Y Y) = \left(\frac{\hat{\alpha}_s}{6\pi}\frac{v f_{0^{++}}}{f^2(m_h^2-m_{0^{++}}^2)}\right)^2\left(1-\frac{v^2}{f^2}\right)\,\Gamma(h(m_{0^{++}})\to YY)_{\mathrm{SM}}
\end{equation}
where $\Gamma(h(m_{0^{++}})\to YY)_{\mathrm{SM}}$ is the partial width of a SM Higgs with its mass taken to be $m_{0^{++}}$. The factor $(1-v^2/f^2)$ is due to the fact that the $h$ couplings to SM particles are suppressed by $\sqrt{1-v^2/f^2}$ compared to their SM values. From Eq.~\eqref{glueballdecay}, by assuming that $\hat{G}_{0^{++}}$ decays only into SM particles, its lifetime is determined from the lifetime of a light SM Higgs. In the mass range of interest to us, $2m_\tau \lesssim m_G \lesssim m_W$, the following simple scaling approximately holds \cite{Craig:2015pha}
\begin{equation}\label{eq:glueballdecay}
c\tau_{\hat{G}_{0^{++}}}\simeq1\;\mathrm{cm}\left(\frac{5\;\mathrm{GeV}}{\Lambda}\right)^7\left(\frac{f}{1\;\mathrm{TeV}}\right)^4. 
\end{equation}
A more accurate value can be obtained by computing the SM Higgs width with HDECAY \cite{Djouadi:1997yw}. For our benchmark values $f=1\;\mathrm{TeV}$, $\Lambda = 5\;\mathrm{GeV}$ this gives a lifetime of $3\;\mathrm{mm}$.

The $\hat{G}_{0^{-+}}$ is long-lived if the twin leptons are heavy. This however does not pose a problem for cosmology, because the pseudoscalar can annihilate efficiently into the $\hat{G}_{0^{++}}$. For the $\hat{G}_{2^{++}}$, the gauge and Lorentz invariance permit the coupling
\begin{equation}
\hat{h}_{\mu\nu}\left(F^{\mu\alpha}F_{\,\,\,\alpha}^{\nu}-\frac{1}{4}g^{\mu\nu}F^{\alpha\beta}F_{\alpha\beta}\right),
\end{equation}
where the spin-$2$ twin glueball field is denoted by a tensor $\hat{h}_{\mu\nu}$, and $F$ indicates either the SM or twin photon field strength. This coupling can originate from the following dimension-$8$ operator, mediated by loops of $\tilde{d}_3^B$
\begin{equation}
\sim \frac{N_c\,g_s^2\,e^2}{16\pi^2}\frac{1}{\tilde{M}^4}\left(\hat{G}_{\mu\beta}^a \hat{G}^{\,\,\,\beta\, a}_{\nu}-\frac{1}{4}g_{\mu\nu}\hat{G}_{\alpha\beta}^a \hat{G}^{\alpha\beta\,a}\right)\left(F^{\mu\alpha}F_{\,\,\,\alpha}^{\nu}-\frac{1}{4}g^{\mu\nu}F^{\alpha\beta}F_{\alpha\beta}\right),
\end{equation}
after the twin gluonic operator in parentheses is replaced by $f_{2^{++}}\hat{h}_{\mu\nu}$. Assuming the twin photon to be relatively heavy, the dominant decay of $\hat{G}_{2^{++}}$ is into two SM photons. The width can be obtained using the results of Ref.~\cite{Han:1998sg} 
\begin{equation}
\Gamma(\hat{G}_{2^{++}}\to\gamma\gamma) = \frac{N_c^2\alpha^2C_{2^{++}}^2}{640\pi^3}\frac{m_{2^{++}}^9}{\tilde{M}^8}\,,
\end{equation}
and the corresponding decay length is
\begin{eqnarray}
c\tau_{\hat{G}_{2^{++}}}&\simeq&4\,\text{km}\,\left(\frac{5\,\text{GeV}}{\Lambda}\right)^9\left(\frac{\tilde{M}}{1\,\text{TeV}}\right)^8\,.\label{eq:Gmptogamma}
\end{eqnarray}
Since $\tilde{M}\gsim\mathrm{TeV}$ from Eq~(\ref{eq:Mbound}), the decay happens outside the detector. 

Given the above lifetime estimates of the twin glueballs, in the analysis of the exotic quark signals we focus on the displaced signature given by the $\hat{G}_{0^{++}}$, which decays into $b\bar{b}$ through Higgs mixing with $O(\rm cm)$ lifetime. The signal process is
\begin{equation}\label{exoquark signal_glueballs}
pp\;\to\; (\mathcal{T}\to t Z_B)(\overline{\mathcal{T}}\to \bar{t}Z_B)\; \to\; t\bar{t} \;+\; \mathrm{twin\;glueballs}\,,\qquad \hat{G}_{0^{++}}\;\to\; (b\bar{b})_{\mathrm{DV}}. 
\end{equation}
The simulation of the signal is very similar to the twin bottomonium search in Sec.~\ref{sec:bottom}. In particular, we assume that in the $Z_B$ rest frame $n_{\hat{G}}/2$ of the produced twin glueballs move along the $\hat{b}$ direction, and the other $n_{\hat{G}}/2$ move along the $\bar{\hat{b}}$. In addition, we assume that the $Z_B$ energy is equally split among the twin glueballs, which are all assumed to have mass equal to the lightest one, $m_{0^{++}}$. The detector modeling and cuts are identical to those used for the $\hat{\chi}_{b0}\to (b\bar{b})_{\rm DV}$ signal. 

To compute the search reach we also need to estimate the $\hat{G}_{0^{++}}$ fraction among the twin glueballs produced. Unlike the string splitting case, the twin glueballs mainly come from twin gluon emission in scattering, which is dominated by low energy processes. Therefore, it is preferable to produce lighter twin glueball states. Given that $\hat{G}_{0^{++}}$ is significantly lighter than the other twin glueballs, we expect that a significant fraction of the twin glueballs produced will be $\hat{G}_{0^{++}}$. For a conservative estimate, we assume that there is one $\hat{G}_{0^{++}}$ produced in each $Z_B$ decay. The corresponding reach at the $13$ TeV LHC and a future $100$ TeV collider is shown in blue in Fig.~\ref{fig:glueball} (lower curves). On the other hand, if the twin leptons are lighter than half of the mass gap between the $\hat{G}_{0^{++}}$ and the excited states, the $\hat{G}_{0^{-+}}$ and $\hat{G}_{2^{++}}$ will decay via off-shell $Z_B$, and end up producing a $\hat{G}_{0^{++}}$ together with twin leptons. The $0^{++}$ state still decays dominantly into SM fermions, because the Higgs coupling to the twin leptons is suppressed by $v/f$. In this case, all the produced glueballs eventually decay down to the $\hat{G}_{0^{++}}$, enhancing the $(b\bar{b})_{\rm DV}$ signal. This is in contrast to the twin bottomonium case, where the lightest twin hadron $\hat{\eta}_b$ prefers to decay into twin leptons if it is kinematically allowed and therefore light twin leptons are expected to deplete the signals. The collider reach of this optimistic scenario is shown by the upper blue curves in Fig.~\ref{fig:glueball}. We can see that the sensitivity to the exotic quark mass based on the displaced twin glueball decay is better than the stop search channel. At the LHC it reaches $m_{\mathcal{T}}\sim 2\,(2.5)$ TeV with $300\,(3000)$ fb$^{-1}$, whereas for the 100 TeV collider it goes up to $\sim 12\,\mathrm{TeV} \sim 4\pi f$. Notice that if the twin leptons from the twin glueball cascade decays further decay back to SM particles via higher dimensional operators, a combination of twin glueball and twin lepton signals is also possible, although it is more model dependent. The region $8 < m_{\hat{b}} < 17$ GeV, shaded in grey in Fig.~\ref{fig:glueball}, corresponds to the regime where string breaking and scattering happen on comparable time scales. In this region we expect a combination of the bottomonium and glueball signals, but it is difficult to make quantitative predictions. The region shaded in red is instead already excluded by the $8$ TeV Higgs fit, see Eq.~\eqref{eq:invHbound}.

\subsection{Twin Lepton Signals}\label{sec:twintau}
%

We now turn to the signals that arise from the twin leptons, if they can decay back to the SM sector within the detector. The masses of the twin tau and twin neutrino are free parameters in the Fraternal Twin Higgs scenario. The twin neutrino mass can be either Majorana, or Dirac type if one also introduces the right-handed twin neutrino. Since both $SU(2)_B$ and $U(1)_D$ gauge groups are broken in the twin sector, twin tau and twin neutrino are singlets under the unbroken gauge groups at low energies. From the SM point of view, they can behave like sterile neutrinos. One could introduce higher-dimensional operators that couple them to the SM neutrino sectors, for instance,
\beq
{\cal O}_{\hat{\nu} {\rm SM}} \=   \frac{1}{M_1} (H_B^\dagger \ell_{3L}^B)( H_A^\dagger \ell_{3L}^A), \qquad
{\cal O}_{\hat{\tau} {\rm SM}} \=  \frac{\langle \phi \rangle}{M_2}\, \overline{\tau}_{3R}^B H_A^\dagger \ell_{3L}^A ,
\label{eq:mixing tauB} 
\eeq
where $\langle \phi \rangle$ is a spurion breaking the $U(1)_D$ symmetry with charge $-1$. These operators induce mass mixings between the twin leptons and the SM neutrinos after substituting in the Higgs VEVs. Therefore they allow the twin leptons to decay into three SM fermions (as long as the phase space is open), either three leptons or one neutrino plus a pair of quarks.\footnote{With the $U(1)_D$ breaking spurion, one can also write down the operator $\frac{\langle \phi \rangle}{M_3} \overline{\tau}_{3R}^B H_B^\dagger \ell_{3L}^B$ which mixes the twin tau and the twin neutrino. If one is more than three times heavier than the other, the heavier one can also decay into three lighter ones.} 

As already mentioned in the previous two subsections, if the twin leptons are light they can be produced in twin hadron decays. However, making robust predictions for these signatures is difficult, because it requires a detailed modeling of the complex twin hadron decays. On the other hand, the exotic quarks also produce twin leptons through purely weak decays: the $\mathcal{B}$ always decays to $t(W_B \to \hat{\tau}\hat{\nu})$. In the following we concentrate on this process. Since both $\hat{\tau}$ and $\hat{\nu}$ act like sterile neutrinos and they can have different mixings with SM neutrinos and have different decay lengths, we will simply consider one of them which can give us favorable signals and denote it as $\hat{\ell}$. If both can decay to SM inside the detector, that will further enhance our signals and the search reach.

In the presence of a sterile neutrino $x$ beyond the three SM ones, we can generally write the gauge eigenstates as linear combinations of the mass eigenstates
\begin{equation}
\nu_{\alpha L} = U_{\alpha i} \,\nu_{i L}\,,\qquad \alpha = e, \mu, \tau\,,\qquad i = 1,2,3,x\,. 
\end{equation}  
With this parameterization, the decays of $x$ are given by \cite{Barger:1995ty,Picciotto:2004rp,Gorbunov:2007ak} \footnote{We used the fact that the mixing matrix in the active sector is unitary up to subleading corrections, which gives
$$
\sum_{l=1,2,3} \Big| \sum_\beta U^\dagger_{x\beta} U_{\beta l} \Big|^2= \sum_\gamma \left| U_{\gamma x}\right|^2\,\qquad \mathrm{and}\qquad \sum_{l=1,2,3} \left| U_{\beta l}\right|^2 = 1\,.
$$
Also, these formulae assume that $x$ is a Dirac fermion. For a Majorana sterile neutrino the decay width will be twice as large, because lepton number does not need to be conserved.
}
\begin{align}
\Gamma(x \to \nu \nu \bar{\nu}) \,&=\, \Gamma_{\mu x} \,\sum_{\gamma=e,\mu,\tau} \left|U_{\gamma x}\right|^2\,, \nonumber \\
\Gamma(x \to \nu \ell^+_\alpha \ell^-_\alpha) \,&=\,\frac{\Gamma_{\mu x}}{4} \left[(1-4s_w^2 + 8 s_w^4) \, \sum_{\gamma=e,\mu,\tau}\left|U_{\gamma x}\right|^2 + 8 s_w^2 \left|U_{\alpha x}\right|^2\right], \nonumber \\
\Gamma(x \to \nu \ell^+_\beta \ell^-_\alpha) \,&=\, \Gamma_{\mu x} \,\left|U_{\alpha x}\right|^2\,\qquad (\alpha \neq \beta), \nonumber \\
\Gamma(x \to \nu q \bar{q}) \,&=\, N_c\, \Gamma_{\mu x} \left(1-2s_w^2+2s_w^4\right) \,\sum_{\gamma=e,\mu,\tau} \left|U_{\gamma x}\right|^2\, \qquad (q=u,d,c,s), \nonumber \\
\Gamma(x \to \ell^-_\alpha q \bar{q}^{\prime}) \,&=\, 2N_c\, \Gamma_{\mu x} \left|U_{\alpha x}\right|^2\,\qquad (q=u,c;\,q'=d,s), 
\label{eq:taupdecay}
\end{align}
where for convenience we defined $\Gamma_{\mu x} \equiv G_F^2 m_x^5/(192\pi^3)$ (muon lifetime with muon mass replaced by $m_x$), and we assumed $m_{x} \lesssim 2m_b\,$. There is also a rare one-loop decay into a photon and a neutrino, which we will neglect.  To illustrate, we assume that the twin lepton mixes directly only with the SM third generation, i.e., the SM lepton doublet in Eq.~\eqref{eq:mixing tauB} being the third generation one,
\begin{equation}
\begin{pmatrix} \nu_{e L} \\ \nu_{\mu L} \\ \nu_{\tau L} \end{pmatrix} = \begin{pmatrix} 1 & & & \\ & 1 & & \\ & & \cos\theta_\nu & -\sin\theta_\nu \end{pmatrix} \begin{pmatrix} \nu_{1 L} \\ \nu_{2 L} \\ \nu_{3 L} \\ \hat{\ell} \end{pmatrix}\,\qquad \rightarrow \qquad U_{\gamma x} = -\sin\theta_\nu\, \delta_{\gamma \tau}\,.
\label{eq:ttau3}
\end{equation}
Summing over the decay modes in Eq.~\eqref{eq:taupdecay}, we find the total width of $\hat{\ell}$ 
\begin{equation}
\Gamma_{\hat{\ell}} = \frac{G_F^2 m_{\hat{\ell}}^5}{192\pi^3}\left(\frac{51}{4}-7s_w^2 + 12 s_w^4 \right)\sin^2\theta_\nu \approx \left( \frac{\sin \theta_\nu}{10^{-3}}\right)^2  \left( \frac{m_{\hat{\ell}}}{6 \GeV} \right)^5 \, \left( \frac{1}{10\; \mathrm{cm}} \right).
\end{equation}
Therefore, the proper decay length is of order of $10-0.1$ m for $\sin\theta_\nu \sim 10^{-4}-10^{-3}$ (which satisfies all constraints on the mixing angle, see e.g.~Ref.~\cite{Izaguirre:2015pga}) and $m_{\hat{\ell}} \simeq 6 \GeV$. This is a favorable range for displaced decay signals to appear at the LHC and future hadron colliders. 
Compared to the twin hadron production via twin QCD hadronization that we studied in Secs.~\ref{sec:bottom} and \ref{sec:glueball}, twin lepton production in purely (twin) weak decays of the $\mathcal{B}$ is affected by smaller theoretical uncertainties. Therefore in this subsection we perform a slightly more refined collider study of the twin lepton signals than the crude approximations adopted in the twin hadron analyses.

The last four decay modes in \Eq{eq:taupdecay} can lead to various DV signals composed of displaced dileptons or displaced hadronic jets. Given the assumption of \Eq{eq:ttau3}, the branching ratios of the twin lepton are independent of $\theta_\nu$. We find BR($\hat{\ell} \to \ell^+ \ell^{(\prime)-} + \nu) \simeq 2.1\%$ for dileptons ($\ell,\ell' = e,\mu$) without including tau leptonic decays. The latter contribute 
an additional $6.1\%$, giving a total branching ratio into leptons of $8.2\%$. The branching ratio into quarks is larger, equal to $67\%$ for $m_{\hat{\ell}} \lesssim 2m_b\,$. Here we do not include the hadronic tau decay, because hadronic taus typically do not produce enough tracks to be detected as DVs. In any case, including them does not increase the hadronic branching ratio significantly. The remaining decay modes involve either a mixture of leptons and hadronic taus or pure neutrinos, which we neglect in our collider study. The processes of interest are then
\begin{equation}\label{exoquark signal_mesons B}
pp\,\to\, (\mathcal{B}\to t W_B)(\overline{\mathcal{B}}\to \bar{t}W_B)\, \to\; t\bar{t} \,+\, \mathrm{twin\;leptons}\,,\quad \hat{\ell}\,\to\,(\ell^+\ell^-)_{\rm DV}\mbox{ or } \hat{\ell}\;\to\; (q\bar{q}^{(\prime)})_{\mathrm{DV}}\,.
\end{equation}
To estimate the prospects for these signals we adopt currently available LHC search strategies and DV reconstruction efficiencies, referring 
to several analyses that as a whole can cover a large range of twin lepton lifetimes. For leptonic $\hat{\ell}$ decays, in addition to the ID dilepton DV search by CMS \cite{CMS:2014hka}, to which we already referred in the $\hat{\Upsilon}$ analysis, we include the ATLAS displaced lepton jets search in the HCAL and MS \cite{Aad:2014yea}. The ID dilepton DV search targets long-lived decays into a dilepton pair in the ID. While only $ee$ and $\mu\mu$ vertices were searched for in $8$ TeV data, the $e\mu$ combination can be produced by the twin lepton decay and it is therefore included. Displaced ID lepton tracks must have $d_0 > 0.2$ mm, where the impact parameter $d_0$ is the distance of closest approach to the primary vertex. We also require an angular separation of $\Delta R > 0.2$ among the leptons, as imposed in Ref.~\cite{CMS:2014hka} for $\mu\mu$ pairs. This requirement may be relaxed by some amount for electron DVs. However, as discussed later, this would not affect our results significantly and therefore we conservatively and simply impose $\Delta R>0.2$ for all dilepton combinations. For same-flavor dileptons, an invariant mass cut $m_{\ell \ell} > 15$ GeV is additionally imposed, to avoid contamination from SM quarkonium decays. The displaced lepton jet search looks instead for displaced collimated ($\Delta R<0.5$) lepton pairs in the HCAL or MS, composed of $ee$ or $\mu \mu$, respectively. For hadronic $\hat{\ell}$ decays we consider the ATLAS hadronic DV searches in the ID, HCAL and MS \cite{Aad:2015uaa,Aad:2015asa}, to which we already made reference for the $\hat{\chi}_{0b}$ and $\hat{G}_{0^{++}}$ analyses. To reject the backgrounds, in the ID hadronic search a minimum number of tracks with relatively large $d_0 > 10$ mm are required.

In addition to the displaced decay products, the signals feature prompt and hard objects stemming from the $t\bar{t}$ pair. As we already mentioned, the combination of a displaced signature with prompt hard objects, in addition to guaranteeing efficient triggering, is potentially free from backgrounds, because the main processes that can fake displaced signals (such as accidental track crossings, non-prompt hadronic decays and photon conversions) are normally not accompanied by hard prompt objects. We require $p_T>50$ GeV and $|\eta|<2.5$ for at least one top quark, which essentially ensures the detection of some hard prompt object(s) from the top decays. 

A realistic study of the signals in Eq.~\eqref{exoquark signal_mesons B} would require dedicated simulations of detector performance, hadronic showering and reconstruction algorithms. Instead, to glimpse search prospects, we simplify the study by using parton-level event samples; applying simplified cuts on partons that ensure DVs can be properly reconstructed; and including the DV reconstruction efficiencies, which are given in the experimental papers as a function of the decay length. The parton-level cuts, as well as the experimental references where the efficiencies can be found, are collected in Table~\ref{tab:dileptonDV-param} for dilepton DVs and Table~\ref{tab:dijetDV-param} for hadronic DVs. Typical DV reconstruction efficiencies are $10\%, 35\%$ and $15\%$ for hadronic DV in the ID, leptonic DV in the ID, and both types of DVs in MS+HCAL, respectively. We use the same cuts and efficiencies for the 13 TeV LHC and a future 100 TeV collider. The signal rate is then obtained multiplying the cross section for $\hat{\ell}$ production, the probability for it to decay in the relevant parts of the detector, the cut efficiencies and the DV reconstruction efficiency. At least one DV is required to be reconstructed in each event.

\begin{table}[t] \centering
\begin{tabular}{c | c | c | c| c | c| c}
detector & $e/\mu$ & $r_{\rm DV}(\hat{\ell})$ range & $\Delta R(\ell, \ell)$ & $p_T(\ell), |\eta(\ell)|$ & $d_0(\ell), m_{\ell \ell}$ & DV eff. \\
\hline
ID  & all & (1, 50) cm & $ \geq 0.2$ & 30 GeV, 2 & $0.2\,$mm, $15\,$GeV & \cite{CMS:2014hka,CMS:2014hka2}\\
MS & $ \mu\mu$ & (0.5, 4) m & $ \leq 0.5$ & 30 GeV, 2 & --  & fig.~6b of \cite{Aad:2014yea} \\
HCAL & $ ee$ & (1, 3.5) m & $\leq 0.5$ & 30 GeV, 2 &  --  &  fig.~7b of \cite{Aad:2014yea} \\
\end{tabular}
\caption{Dilepton DV parameters used in our parton-level study. If the leptons satisfy these cuts, we assume that the DV is reconstructable with the efficiency taken from the reference in the last column. In addition, $p_T>50$ GeV and $|\eta|<2.5$ is imposed on at least one top quark.}
\label{tab:dileptonDV-param}\end{table}
\begin{table}[t] \centering
\begin{tabular}{ c | c | c | c| c | c}
detector & $r_{\rm DV}(\hat{\ell})$ range  & $\Delta R(j,j)$ & $p_T(j), |\eta(j)|$ & $d_0(j)$ & DV eff. \\
\hline
ID  &  (4, 28) cm & -- & 30 GeV, 2 &  $10\,$mm&  fig.~6 of \cite{Aad:2015uaa} \\
MS &  (4, 8) m & -- & 30 GeV, 2 & -- &  fig.~7 of \cite{Aad:2015uaa} \\
HCAL &  (1.9, 3.5) m  & -- & 30 GeV, 2 & -- &  fig.~1a of \cite{Aad:2015asa} \\
\end{tabular}
\caption{Hadronic DV parameters used in our parton-level study. We require at least one jet to pass these cuts, because a single jet alone can leave multiple tracks and be reconstructable as a DV. In addition, $p_T>50$ GeV and $|\eta|<2.5$ is imposed on at least one top quark.}
\label{tab:dijetDV-param}\end{table}
\begin{figure}[t] 
\includegraphics[width=0.48\textwidth]{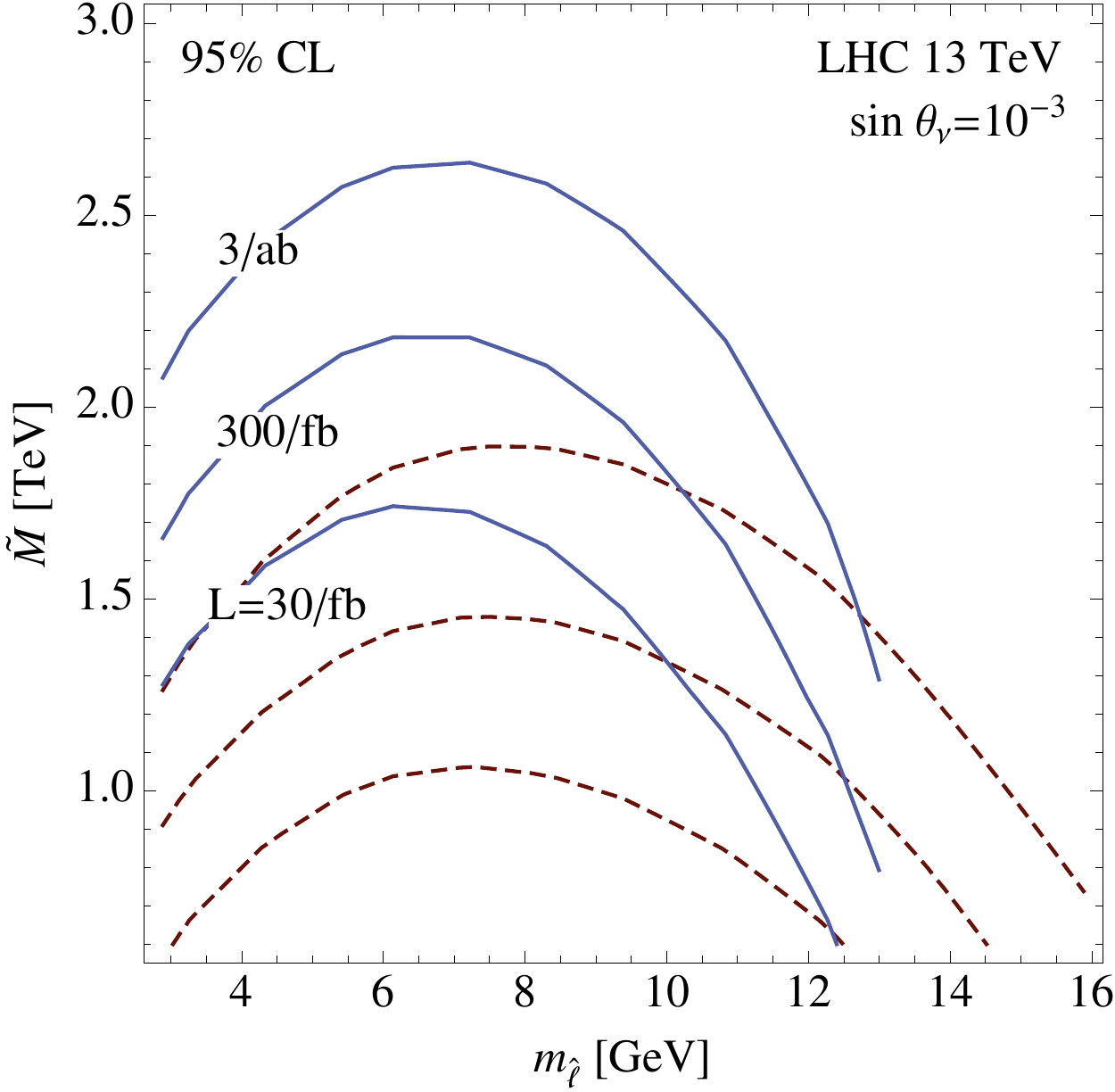}\hspace{3mm}
\includegraphics[width=0.48\textwidth]{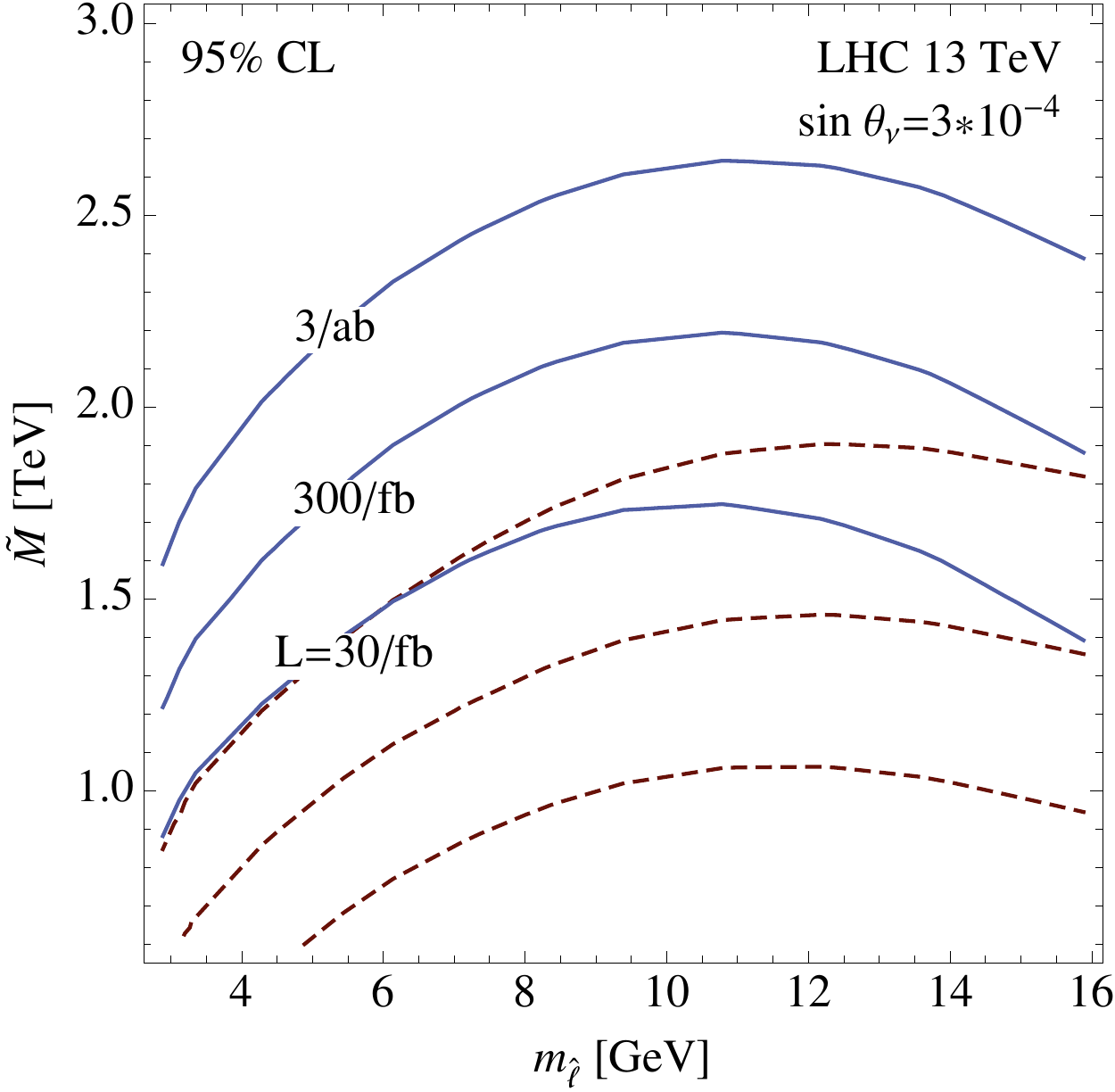}\vspace{2mm}\\
\includegraphics[width=0.48\textwidth]{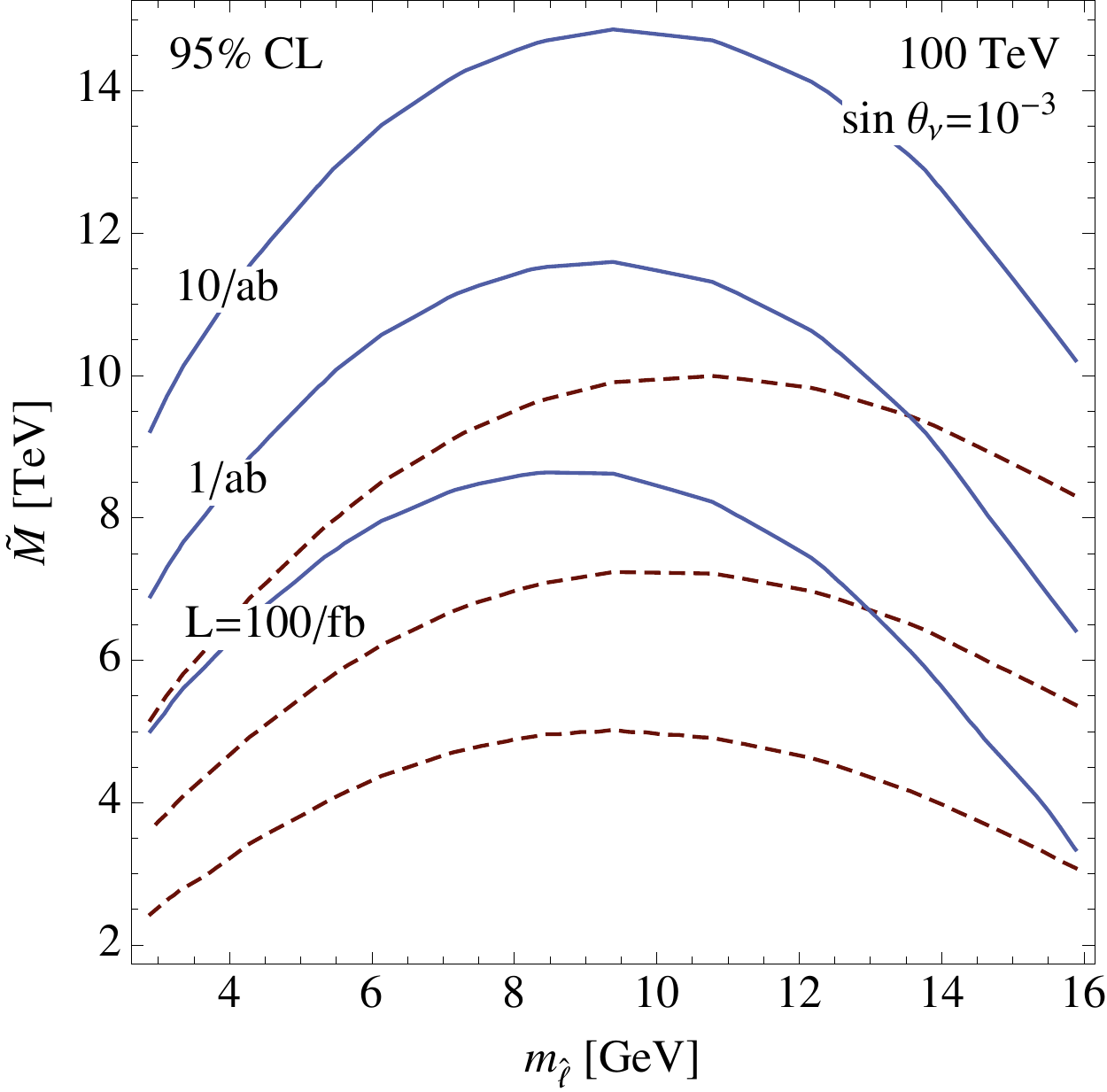}\hspace{3mm}
\includegraphics[width=0.48\textwidth]{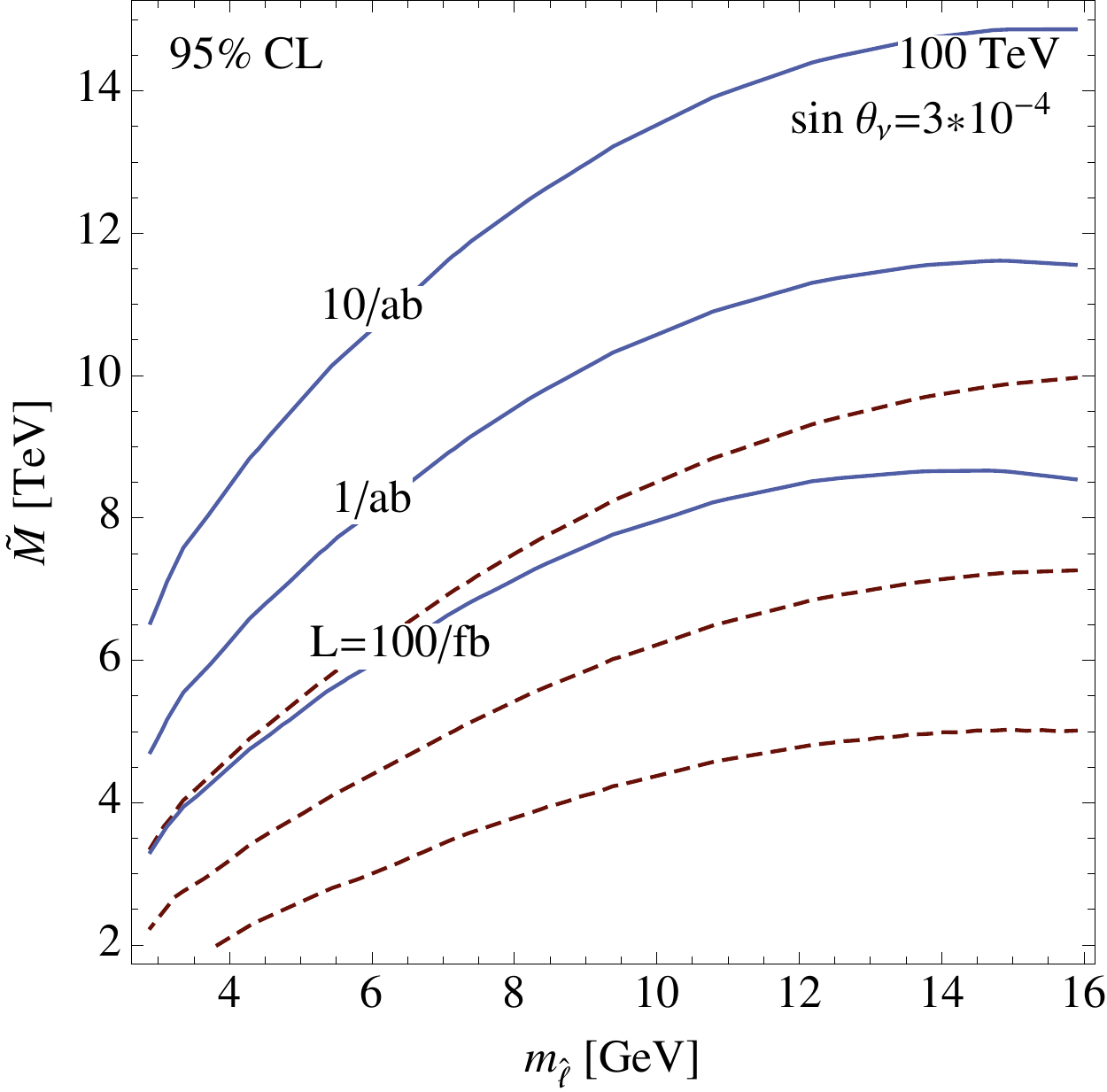}
\caption{Projected exclusion reach on the mass of the exotic quarks from the twin lepton DV signal, at the $13$ TeV LHC (top row) and 100 TeV collider (bottom row), assuming $\sin \theta_\nu = 10^{-3}$ (left column) and $3\times 10^{-4}$ (right column). Blue solid lines correspond to hadronic DV searches and brown dashed lines to dilepton DV searches. DV searches in all parts of detector are combined, but the ID hardly contributes. The scale $f$ does not directly affect the results, as discussed in the text. 
Selection cuts are listed in Table~\ref{tab:dileptonDV-param} for dilepton DV and Table~\ref{tab:dijetDV-param} for hadronic DV, supplemented by mild cuts on at least one top quark: $p_T(t)>50$ GeV and $|\eta(t)|<2.5$.}
\label{fig:DV-res}
\end{figure}
Under the assumption of zero background, we can exclude a signal hypothesis if more than $3$ events are expected and none observed. 
In Fig.~\ref{fig:DV-res} we show the resulting exclusion reach on $\tilde{M}$ at the 13 TeV LHC and at the 100 TeV collider, as a function of the $\hat{\ell}$ mass. At the 13 TeV LHC, the maximum reach on $\tilde{M}$ is obtained for $c\tau_{\hat{\ell}} \sim 40$ mm, which for $\sin \theta_\nu=10^{-3}$ (top left panel in Fig.~\ref{fig:DV-res}) corresponds to $m_{\hat{\ell}} \sim 7-8$ GeV. For a heavier (lighter) twin lepton, the decay length becomes too short (long), and the sensitivity drops. For smaller $\sin \theta_\nu$ (right top panel) the optimal $m_{\hat{\ell}}$ increases, to compensate the suppression from the smaller mixing angle. At $100$ TeV (lower panels) the best sensitivity is obtained for shorter lifetime, $c\tau_{\hat{\ell}} \sim 10$ mm, due to the larger boost. In the optimal case, the 13 TeV LHC with 300 fb$^{-1}$ can probe up to $\tilde{M} \sim $ 2.2 TeV, while a 100 TeV collider with 1 ab$^{-1}$ will be able to reach $\sim 11$ TeV. Hadronic DV searches give stronger bounds than leptonic DV searches, mainly because of the larger hadronic branching ratio of $\hat{\ell}$. If the twin lepton lifetime is much shorter than the optimal value for the DV signal, a search for multiple prompt jets and leptons may be effective. For a much longer lifetime, we are left with the stop-like signal discussed in Sec.~\ref{sec:stop}, which was shown to be sensitive to $\tilde{M}\sim 8$ TeV at the 100 TeV collider with $1$ ab$^{-1}$.

In Fig.~\ref{fig:DV-res} DV signals in the ID, HCAL and MS are all added, but the ID contribution is very suppressed, because the large boost of $\hat{\ell}$ implies small $\Delta R \ll {\cal O}(0.1)$ and small $d_0 \ll {\cal O}(10)$mm. In addition, in the leptonic ID search the cut $m_{\ell\ell}>15$ GeV removes the same-flavor dilepton decays of $\hat{\ell}$, leaving only the smaller $e\mu$ component. Although to be conservative in the ID DV analyses we imposed either $\Delta R \geq 0.2$ or $d_0 >10$ mm, it might be possible to relax those requirements to improve the searches. Indeed, the track resolution in hadronic jets is of $\Delta R \sim {\cal O}(0.01)$ or better at both ATLAS \cite{ATLAS:1999uwa} and CMS \cite{Bayatian:2006zz}, and some hadronic DV searches do use smaller $d_0 > 0.5$ mm \cite{CMS:2014wda}. If such improvements could be implemented in ID DV searches, the sensitivity to smaller decay lengths in our model would be significantly improved.

Finally, we note that the symmetry breaking scale $f$ does not enter Fig.~\ref{fig:DV-res}, because the twin lepton signal does not depend directly on it, but only on the mass $m_{\hat{\ell}}$ and the mixing angle $\theta_\nu$. 

\section{Discussion and Conclusions}
\label{sec:conclusions}
In this paper we considered the phenomenology of non-supersymmetric UV completions of the Twin Higgs mechanism, focusing on the exotic fermions that carry both SM and twin gauge charges. Since these states regularize the logarithmic divergences in the top contribution to the Higgs potential, their masses are expected to lie in the $1$--$10$ TeV range. Some of them are SM-colored and thus provide a new portal to access the twin sector from above, through their prompt decay into a top quark and twin particles. If all the twin particles leave no trace in the detector, we are left with the stop-like irreducible signal of $t\bar{t}$ plus missing transverse energy. However, some of the twin particles may decay back to the SM with long lifetimes, leading to spectacular combinations of prompt and displaced signatures that are virtually free from SM backgrounds. Depending on the twin sector parameters, the displaced signals are dominated by twin hadrons (bottomonia or glueballs) or twin leptons. The twin hadrons decay to SM particles either through the Higgs portal, or the gauge kinetic mixing induced by loops of the exotic fermions. In the latter case, collider experiments and cosmological observations provide complementary constraints. The twin leptons instead can couple to the SM via higher-dimensional operators. Our results show that, depending on the details of the twin sector, the exotic quarks can be probed up to $\sim 2$ TeV at LHC Run 2, and beyond 10 TeV, the upper limit of the UV cutoff of a natural low energy theory, at a future $100$ TeV collider. This further supports the case for a future hadron collider as discovery machine for neutrally natural theories of the weak scale \cite{Curtin:2015bka}.     

Our discussion was set in the Fraternal Twin Higgs model, and most of the phenomenological results presented in the paper were based on the benchmark values $f = 1\;\mathrm{TeV}$ and $\Lambda = 5\;\mathrm{GeV}$, which are motivated by naturalness arguments. To understand the degree of generality of the signals we studied, we now wish to discuss how they would be affected by relaxing each of these assumptions. We start by considering different $f$ and $\Lambda$ values within the Fraternal Twin framework, then followed by some discussion of departures from this model. 

The choice of $f=1\;\mathrm{TeV}$ guarantees that no deviation in the Higgs couplings would be observed at the LHC, thus further strengthening the relevance of direct signatures, while keeping only a mild electroweak tuning of $\sim 10\%$. Larger values of $f$ can be considered at the expenses of more tuning. The masses of the twin gauge bosons scale with $f$ so they become heavy in this case, resulting in a larger twin hadron multiplicity in the exotic quark decays. At the same time, the Higgs mediation between the SM and twin sectors is suppressed, leading to longer lifetimes for the $0^{++}$ twin bottomonium and twin glueball ($\propto f^4$ for a fixed hadron mass), which can enhance the displaced signals. For very large values of $f\gtrsim 5\;\mathrm{TeV}$ the twin hadrons decay outside of the detector and we are left with the irreducible signal, but at this point the Twin Higgs model already loses its appeal as a natural theory of the electroweak symmetry breaking.  

The $\Lambda = 5$ GeV was picked as the benchmark because it corresponds to the equality at high scale of the SM and twin QCD couplings, which ensures that the dominant two-loop quadratic corrections to the Higgs mass are also canceled exactly. The tuning would still be acceptable for a small enough difference between the two couplings, which corresponds to $1 < \Lambda < 20$ GeV. The value of $\Lambda$ determines the separation between the regions where the twin bottomonium ($m_{\hat{b}}\lesssim \Lambda$) or twin glueball ($m_{\hat{b}}\gg \Lambda$) signals dominate. Increasing $\Lambda$ makes the twin color string easier to break, therefore the string fragmentation picture applies up to larger $m_{\hat{b}}$. At the same time the twin meson mass increases, leading to a smaller multiplicity but also to larger transverse momenta of the individual hadrons, with a partial compensation of the two effects. The decay lengths of the twin hadrons decrease rapidly with increasing $\Lambda$, requiring the reconstruction of displaced vertices down to smaller distances from the beamline. For $\Lambda\gtrsim 15$ GeV, the lifetime of the $0^{++}$ bottomonium is shorter than $0.1$ mm, and the decay effectively becomes prompt. Conversely, for smaller $\Lambda$ the string de-excitation through glueball emission dominates over a larger region of $m_{\hat{b}}$, and the twin hadron lifetimes increase. For $\Lambda\lesssim 1.5$ GeV the $0^{++}$ glueball decays outside of the detector, and we are left with the irreducible missing transverse energy signal.

Next we consider departures from the Fraternal Twin model. As long as the twin leptons and photon remain heavy and there is no light Goldstone bosons from spontaneous chiral symmetry breaking, most of our results can still apply. In this case, the lightest twin quark would be identified with the $\hat{b}$ in our discussion. For example, in the model of Ref.~\cite{KnapenETALtoappear} with one generation of vector-like twin quarks, the theory is anomaly-free without the presence of the twin leptons. In this case, the lightest twin top mass eigenstate can play a role similar to the $\hat{b}$, and displaced vertex signals can arise from twin toponia or glueballs. On the other hand, if a complete mirror copy of the SM is considered, at the bottom of the twin spectrum we expect to find the twin pions, the lightest twin baryons and the twin leptons to be long-lived. Depending on the parameters, these may or may not decay back to the SM on collider time scales. In the most pessimistic scenario where all of them are too long lived, we are again left with the irreducible signal.

Finally, we conclude by commenting on the other states that are expected to accompany the exotic quarks in the UV completion discussed in this paper. The exotic fermion multiplet $\tilde{q}_3^B$, whose $Z_2$-symmetric mass is $\sim \tilde{M}$, carries electroweak charge and twin color, and is therefore pair-produced with a suppressed rate at hadron colliders. It decays primarily into a twin top plus a SM $W,Z$. On the other hand, one also expects to find exotic gauge bosons that restore the full $SU(4)$ invariance of the Higgs kinetic term in the UV. Similarly to the exotic fermions, the masses of these vectors regularize the logarithmic divergences in the gauge contribution to the Higgs potential. In particular, the off-diagonal exotic gauge bosons $\tilde{W}$ couple to the exotic fermions with interactions of the form $\bar{\tilde{q}}_{3}^{A,B} \slashed{\tilde{W}}q_{3}^{A,B}$. Therefore, depending on the relative masses, the exotic quarks may decay into the $\tilde{W}$, or vice versa. 
In addition, there may be exotic gauge bosons which complete the $SU(6)$ multiplet in the UV theory. Although they are not strictly required for the finiteness of the Higgs potential at one loop, they appear in some UV-complete models. The off-diagonal ones $\tilde{G}$ carry both SM and twin color, and their couplings to exotic fermions have the form $\bar{\tilde{q}}_{3}^{B,A} \slashed{\tilde{G}}q_{3}^{A,B}$. Since $\tilde{G}$ can be produced by the strong interaction, its decays may provide a promising production mechanism for the elusive $\tilde{q}_{3}^B$, if it is kinematically allowed. The study of this phenomenology is left for future work.

\section*{Acknowledgments}
We would like to thank Zackaria Chacko, Roberto Contino, David Curtin, Andy Haas, Simon Knapen, Markus Luty, Agostino Patella, Matt Strassler, and Jesse Thaler for useful discussion. HC thanks for hospitality Academia Sinica and GGI, where part of the work was done, and the INFN for partial support. ES and YT are grateful to the MIAPP and GGI for hospitality and partial support during the completion of this project. YT also thanks KIAS and CETUP* for hospitality and partial support. HC, ES, and YT are supported in part by the US Department of Energy grant DE-SC-000999. SJ is supported by the US Department of Energy under contract DE-AC02-76SF00515. YT is supported in part by the National Science Foundation Grant PHY-1315155, and by the Maryland Center for Fundamental Physics.

\appendix
\section{Twin Hadron Multiplicity from the $Z_B$ Decay}\label{app:multiplicity}
As discussed in Sec.~\ref{sec:hiddenpheno}, depending on the relative sizes of $m_{\hat{b}}$ and $\Lambda$, the highly excited $\hat{b}\bar{\hat{b}}$ bound state from the $Z_B$ decay loses its energy either by glueball emission or string breaking. In this appendix, we use a simplified string model to estimate the number of twin hadrons produced by the $Z_B$ decay. 
\subsection{Meson Decay via String Breaking}
When the $\hat{b}\bar{\hat{b}}$ pair is produced from the longitudinal $Z_B$ decay, a string is formed between them. The twin quarks are in a bound state with zero orbital angular momentum and total energy $E_0+2m_{\hat{b}}\equiv m_{Z_B}$. To describe the string breaking we make the simplified assumption that the string breaks at the center through production of a new $\hat{b}\bar{\hat{b}}$ pair. In the $Z_B$ rest frame the twin quarks being pulled out from the vacuum tend to be at rest due to the exponentially suppressed probability $\propto\exp(-E_{\hat{b}}^2/\Lambda^2)$. If the breaking happens when the $\hat{b}$'s at the ends of the string oscillate to the distance $\epsilon\,r_{max}/2$ from the center, with $r_{max}\times 3\Lambda^2\simeq E_0$, the momentum of each twin quark $|\vec{p}_{\hat{b}}|$ right before the string breaking can be solved from the following expression ($E_0+2m_{\hat{b}}=m_{Z_B}$ in the first splitting, and breaking the string requires $m_{\hat{b}}<\epsilon E_0/2$)
\begin{equation}
E_0+2m_{\hat{b}}-\epsilon E_0=2\sqrt{m_{\hat{b}}^2+|\vec{p}_{\hat{b}}|^2}.
\end{equation}
In the absence of external forces, the newly produced strings carry momentum $|\vec{p}_{\hat{b}}|$. Using the energy conservation again for the two shorter strings, we have
\begin{equation}
m_{Z_B}=2\sqrt{m_{inv}^2+|\vec{p}_{\hat{b}}|^2}.
\end{equation}
The invariant mass $m_{inv}$ is a combination of the potential and kinetic energy inside the string, which is a frame-invariant quantity. Once we move to the center of mass frame of the new string and assume the breaking happens at any point between $2m_{\hat{b}}/E_0<\epsilon<1$ with equal probability, the average of the internal energy becomes
\begin{equation}\label{eq:breaking} 
\langle E_1\rangle= \langle m_{inv}\rangle-2m_{\hat{b}},\quad\,\langle m_{inv}\rangle=\frac{E_0}{2(1-2m_{\hat{b}}/E_0)}\int_{2m_{\hat{b}}/E_0}^{1}d\epsilon\sqrt{4\tau^2+4\tau\epsilon+(2-\epsilon)\epsilon},\quad\tau\equiv\frac{m_{\hat{b}}}{E_0}\,.
\end{equation}
For the next breaking, the internal energy of the new strings is a recursion of the energy $E_1\to E_2$, $E_0\to E_1$ in the same expression. For $f=1$ TeV, $m_{\hat{b}}=8$ GeV, and $E_0+2m_{\hat{b}}=m_{Z_B}=360$ GeV, we have
\begin{equation}\label{eq:stringsplit}
E_1=129\,\text{GeV},\qquad E_2=44\,\text{GeV},\qquad E_3=11\,\text{GeV}.
\end{equation}

Since the meson energy after the third breaking $2m_{\hat{b}}+E_3$ is less than twice the expected ground state meson mass [$\simeq 2(m_{\hat{b}}+\Lambda)$], the breaking process will stop when the internal energy is $\simeq E_3$ after the original string breaks into $\sim 8$ pieces. One more breaking will happen for $m_{\hat{b}}\leq0.5$ GeV in this simplified picture. Of course, the breaking does not need to occur symmetrically and we expect that the number of mesons produced increases smoothly as $m_{\hat{b}}$ decreases. 
The average final meson mass is about $3(m_{\hat{b}}+\Lambda)$, in between one and two times of the ground state meson mass. In the simplified model analysis, we usually find that the energies ending up in the masses and the kinetic energies of the final mesons are comparable. Therefore we adopt the simple formula
\begin{equation}
n_{\hat{B}}=\frac{m_{Z_B}/2}{3(m_{\hat{b}}+\Lambda)}
\end{equation}
as our estimate of the final meson multiplicity in the case of string breaking. For the convenience of our collider study we also round it down to an even number.

\subsection{Meson Decay via Glueball Emission}
In the glueball emission case, we use the probability function in Eq.~(\ref{eq:radprob}) to estimate the energy of each gluon emission. Starting from an initial internal energy $E_0\equiv m_{Z_B}-2m_{\hat{b}}$ of the $\hat{b}\bar{\hat{b}}$ pair, the average fraction of it that is carried by the first emitted gluon can be derived as
\begin{equation}
\langle r\rangle=\int_{0}^1dr\,r\,\frac{dP(r)}{dr}=1-\frac{\pi}{4\sqrt{\hat{\alpha}_s}}\,e^{\frac{\pi}{16\hat{\alpha}_s}}\,\text{Erfc}\left[\frac{\sqrt{\pi}}{4\sqrt{\hat{\alpha}_s}}\right].
\end{equation}
When the energy of the emitted gluon is within $35-100$ GeV, we have the ratio to be at least $\langle r\rangle\gsim 0.24$. This means that after each scattering $\gsim 24\%$ of the initial energy $E_{\rm in}$ is emitted into a glueball, and the energy left in the $\hat{b}\bar{\hat{b}}$ system is given by $(1-r)\,E_{\rm in}\,$. 
The process repeats until the average energy emitted in the next scattering would be below the mass of the lightest glueball, $m_{{0^{++}}}\simeq 6.8\,\Lambda$. The remaining energy stored in the string is then radiated through the emission of soft glueballs. Furthermore, if $m_{\hat{b}}$ is sufficiently large annihilation of the de-excited string into two glueballs becomes kinematically accessible. The combination of these three processes gives the expected number of emitted glueballs. For the benchmark point $(\Lambda,\,m_{Z_B})=(5,\,360)$ GeV, we find $6\,\hat{G}+1\;\mathrm{bottomonium}$ for $m_{\hat{b}}<29$ GeV, $8\,\hat{G}$ for $29 < m_{\hat{b}}< 52$ GeV, $6\,\hat{G}$ for $52 < m_{\hat{b}}< 106$ GeV, $4\,\hat{G}$ for $106 < m_{\hat{b}}< 141$ GeV, $2\,\hat{G}$ for $m_{\hat{b}}> 141$ GeV. In these estimates, we made the conservative assumption that, when kinematically allowed, the bottomonium annihilation always produces two glueballs. (This is likely to be an underestimate at large $m_{\hat{b}}$.) We also rounded down to an even number the total multiplicity of emitted glueballs, to simplify the description of the  kinematics of the $Z_B$ decay (see Sec.~\ref{sec:glueball}).

\section{The Bottomonium Decay}\label{app:bottomonium}
In this appendix we present more details about the bottomonium decay. We first compute the decay rates using the \emph{quirky} picture introduced in Ref.~\cite{Kang:2008ea}. Then, as a test of our results, we apply the same estimate to compute the decay rates of SM mesons, finding satisfying agreement with experimental data. 
\subsection{A Quirky Description}
For $m_{\hat{b}}\gg\Lambda$, we can compute the decay rate of the excited bottomonium using the quirky picture. The decay rate of the bottomonium $\hat{B}$ into a pair of particle $X$ can be estimated by
\begin{equation}\label{eq:annicross}
\Gamma(\hat{B}\to XX)=\text{Prob}(r\leq r_0)\times\frac{\sigma_{\hat{b}\bar{\hat{b}}\to XX}\,v_{rel}}{\frac{4\pi}{3}r_0^3}.
\end{equation}
The radius $r$ is the distance between the two $\hat{b}$ twin quarks. The Compton wavelength of the $\hat{b}$, $r_0\sim m_{\hat{b}}^{-1}$, sets the maximum distance at which the two $\hat{b}$'s can annihilate. We denote by Prob$(r\leq r_0)$ the probability of having two $\hat{b}$'s within a relative distance $r_0$, and $\sigma_{\hat{b}\bar{\hat{b}}\to XX}$ is the annihilation cross section of $\hat{b}\bar{\hat{b}}\to XX$. The cross section times the relative velocity in the center of mass frame $v_{rel}$, divided by the volume inside $r_0$, gives the annihilation rate of the process. To estimate Prob$(r\leq r_0)$, we follow the discussion in Ref.~\cite{Kang:2008ea}. Since the twin quarks interact very weakly with the SM nuclei in the detector, the string remains in a quantum state, and we can use the WKB approximation:
\begin{equation}
\text{Prob}(r\leq r_0)=\int_0^{r_0}dr'\,|y(r')|^2,\qquad y(r)\simeq\frac{C}{\sqrt{k(r)}}\sin\left[\int_0^r dr'\,k(r')\right]\,\theta(r_{max}-r),
\end{equation}
where $y(r)$ is the wave function of the $\hat{b}$ at the end of the string relative to the other. The ``local momentum'' $k(r)$ is related to the center of mass energy by 
\begin{equation}
k(r)=\frac{\sqrt{2\mu_{\hat{b}}}}{\hbar}\sqrt{K-V(r)},\qquad K=E-2\mu_{\hat{b}},\quad\mu_{\hat{b}}=m_{\hat{b}}/2.
\end{equation}
The WKB wavefunction serves as a solution of Schr\"odinger's equation $y''(r)=-k^2(r)y(r)$ when $k'(r)\ll k(r)^2$, which holds for a linear potential $V\sim \Lambda^2\,r$ with $\Lambda<K$, where $K$ is the kinetic energy. $r_{max}$ is the classical turning point of the string and $C$ is a normalization constant. In the regime of interest, the kinetic energy is given by $K\sim m_{\hat{b}}\gg V(r)\lsim\Lambda^2/m_{\hat{b}}$ for $r\lesssim r_0$. Thus we can carry out the integration and obtain
\begin{equation}\label{eq:prob}
\text{Prob}(r\leq r_0)=\int_0^{r_0}dr'\,\frac{|C|^2}{m_{\hat{b}}}\sin^2\left(m_{\hat{b}}\,r'\right) \simeq \frac{4}{5}\,\left(\frac{\Lambda}{m_{\hat{b}}}\right)^2.
\end{equation}
Here we have used the normalization
\begin{equation}
|C|^2=\frac{m_{\hat{b}}}{T},\qquad T=\frac{\sqrt{2}\,m_{\hat{b}}}{\sigma},
\end{equation}
and approximated the string potential to be \cite{Bali:2000vr} 
\begin{equation}
V(r)= 3 \Lambda^2 \,r\, .
\end{equation}
The result in Eq.~\eqref{eq:prob} has a simple classical interpretation \cite{Kang:2008ea}: since the maximum string length is $r_{max}\sim m_{\hat{b}}/(3\Lambda^2)$ and the size of the scattering region is $r_0\sim m_{\hat{b}}^{-1}$, the probability for the scattering to happen is $r_0/r_{max}$ (times a numerical factor). The decay width in Eq.~(\ref{eq:annicross}) then takes the expression
\begin{equation}\label{quirk decay}
\Gamma(\hat{B}\to XX)\simeq \frac{3}{5\pi}\Lambda^2\,m_{\hat{b}}\, \sigma_{\hat{b}\bar{\hat{b}}\to XX}\,v_{rel}\,\qquad\qquad (m_{\hat{b}}\gg \Lambda).
\end{equation}
In the opposite regime, $m_{\hat{b}}\ll \Lambda$, no suppression from the wavefunction overlap is expected. Therefore we simply extrapolate Eq.~\eqref{quirk decay} using na\"ive dimensional analysis,
\begin{equation}\label{NDA decay}
\Gamma(\hat{B}\to XX)\simeq \frac{3}{5\pi}\Lambda^3\, \sigma_{\hat{b}\bar{\hat{b}}\to XX}\,v_{rel}\,\;\;\;\;\;\;\qquad\qquad (m_{\hat{b}}\ll \Lambda).
\end{equation}

The $0^{++}$ state $\hat{\chi}_{b0}$ dominantly decays via $\hat{b}\bar{\hat{b}}\to b\bar{b}$ through $h$ exchange. The $h\bar{\hat{b}}\hat{b}$ coupling is obtained from the twin bottom Yukawa $-y_{\hat{b}}\tilde{H}_B^\dagger \overline{d}_{3R}^B q_{3L}^B + \mathrm{h.c.}$ and reads $(y_{\hat{b}}/\sqrt{2})(v/f)h\bar{\hat{b}}\hat{b}$. The annihilation cross section is calculated to be (with no average over initial polarizations)
\begin{equation} \label{b-onium decay}
\sigma_{\hat{b}\bar{\hat{b}}\to b\bar{b}}=N_c \frac{(v/f)^2\,y_{\hat{b}}^2\,y_{b}^2}{8\pi\,v_{rel}}\frac{s-4m_{\hat{b}}^2}{(s-m_h^2)^2}\left(1-\frac{4m_b^2}{s}\right)^{3/2},
\end{equation}
where $\sqrt{s}$ is the center-of-mass energy. Therefore the decay width is, in the regime $m_{\hat{b}}\gg \Lambda$,
\begin{equation}
\Gamma(\hat{\chi}_{b0} \to b\bar{b})= \frac{3N_c}{40\pi^2}\Lambda^2 m_{\hat{b}}\left(\frac{v}{f}\right)^2\,y_{\hat{b}}^2\,y_{b}^2 \frac{s-4m_{\hat{b}}^2}{(s-m_h^2)^2}\left(1-\frac{4m_b^2}{s}\right)^{3/2},
\end{equation}
where we made use of Eq.~\eqref{quirk decay}. Neglecting for simplicity the decay into $\tau^{+}\tau^{-}$, which is suppressed by an order of magnitude, we thus obtain the proper decay length given in Eq.~(\ref{B0decay_quirk}). For $m_{\hat{b}}\ll \Lambda$ we use instead Eq.~\eqref{NDA decay}, resulting in the expression given in Eq.~\eqref{B0decay_NDA}.

The $1^{--}$ state $\hat{\Upsilon}$ decays to the SM via a kinetic mixing operator $(-\epsilon/2)B_{A \mu\nu}B_B^{\mu\nu}$. We compute its decay width in the limit where the breaking of the twin electric charge is small, $m_{\hat{A}}^2 \ll m_{Z_{B}}^2$, so that twin and SM photon exchange dominates the amplitude. In the opposite regime where the twin photon is much heavier than $gf$, this result misses some $O(1)$ factors, but should otherwise give a reasonable estimate. By considering only the diagram with twin photon $\hat{A}$ and SM photon exchange, we find the cross section for $\hat{b}\bar{\hat{b}}\to f\bar{f}$, where $f$ is a SM fermion (not averaged over initial polarizations)
\begin{equation}
\sigma_{\hat{b}\bar{\hat{b}}\to f\bar{f}} \,v_{rel} = \frac{2c_w^4 e^4}{3\pi} \hat{Q}_{\hat{b}}^2 \epsilon^2   \,\frac{s+2m_{\hat{b}}^2}{(s-m_{\hat{A}}^2)^2}\,N_{c}^f Q_{f}^2\,.
\end{equation}
To obtain the total width we need to sum over the SM fermions that are kinematically accessible, which we assume to be all but the top. This leads to $\sum_{f} N_{c}^f Q_{f}^2 = 20/3$. In the regime $m_{\hat{b}}\gg \Lambda$ we apply Eq.~\eqref{quirk decay}, and obtain (taking into account a factor $(2J+1)^{-1} = 1/3$ from average over the spin states of $\hat{\Upsilon}$)
\begin{equation}
\Gamma_{\hat{\Upsilon}} = \frac{8\, c_w^4 e^4}{81\pi^2}\Lambda^2 m_{\hat{b}}\epsilon^2 \,\frac{s + 2m_{\hat{b}}^2}{(s-m_{\hat{A}}^2)^2}\,,
\end{equation}
which in turn leads to Eq.~\eqref{eq:1--decay_quirk}. In the opposite case $m_{\hat{b}}\ll \Lambda$ we use instead Eq.~\eqref{NDA decay}, finding Eq.~\eqref{eq:1--decay_NDA}.

\subsection{Comparison with Experimental Data on Meson Decays}
To check the validity of our approximation, we apply Eq.~\eqref{quirk decay} to the decay of QCD mesons into $e^+ e^-$. We consider the $c\bar{c}$ vector meson $\psi(4415)$. Its mass $M=4.421\;\mathrm{GeV}$  is close to $3 (m_c + \Lambda_{\rm QCD})$, analogous to the typical twin mesons that we consider. Therefore we expect  $\psi(4415)$ to be closer to our approximation. The cross section for $q\bar{q}\to e^+ e^-$ via photon exchange, not averaged over initial polarizations, reads
\begin{equation}
\sigma_{q\bar{q}\to e^+ e^-} = \frac{32\pi\alpha^2 Q_q^2 Q_e^2}{3 s\, v_{rel}}\left(1+\frac{2m_q^2}{s}-\frac{6m_q^4}{s^2}\right)\, .
\end{equation}  
Using as inputs $\sqrt{s}=M=4.421\;\mathrm{GeV}$, $m_c = 1.2\;\mathrm{GeV}$, $\Lambda_{\rm QCD}=0.250\;\mathrm{GeV}$ and $\alpha = 1/137$, we obtain
\begin{equation}
\Gamma(\psi(4415)\to e^+ e^-) = \frac{3}{5\pi}\Lambda_{\rm QCD}^2\,m_{c}\sigma_{c\bar{c}\to e^+e^-}\,v_{rel} \simeq 0.65 \,(0.58) \times 10^{-6}\, \mathrm{GeV},
\end{equation}
where the number in parentheses is the experimental value \cite{PDG}. The agreement is good. In the light quark limit, applying Eq.~\eqref{NDA decay} to the decay of $\rho(770)$ into $e^+ e^-$ we find roughly
\begin{equation}
\Gamma(\rho(770)\to e^+ e^-) = \frac{3}{5\pi}\Lambda_{\rm QCD}^3\sigma_{u\bar{u}\to e^+e^-}\,v_{rel} = \frac{128\alpha^2\Lambda_{\rm QCD}^3}{45 s}\simeq 4.0\, (6.9)\times 10^{-6}\,\mathrm{GeV} ,
\end{equation}
where the experimental value is in parentheses \cite{PDG}, and we used as the input $\sqrt{s} = M = 0.769\;\mathrm{GeV}$. We see that the agreement is within  a factor $2$ even in the light quark case.

\end{document}